\documentclass[aps,prb,twocolumn,showpacs]{revtex4-1}
\pdfoutput=1
\usepackage{graphicx,epsf}
\usepackage{amsmath}
\usepackage{amssymb}
\usepackage{color}
\usepackage{esint}
\usepackage{wasysym}

\newcommand{\bS}{{\bf S}}
\newcommand{\bR}{{\bf R}}
\newcommand{\bq}{{\bf q}}
\newcommand{\bQ}{{\bf Q}}
\newcommand{\bp}{{\bf p}}
\newcommand{\bk}{{\bf k}}

\newcommand{\taub}{\mbox{\boldmath $\tau $}}

\newcommand{\xdmit}{X[Pd(dmit)$_2$]$_2$}
\newcommand{\pdmit}{EtMe$_3$P[Pd(dmit)$_2$]$_2$}

\definecolor{darkgreen}{rgb}{0,0.8,0}
\definecolor{orange}{cmyk}{0.1,0.2,1,0}
\definecolor{magenta}{cmyk}{0,0.9,0,0}

\begin{document}

\title{
 Triangular-lattice anisotropic dimerized Heisenberg antiferromagnet: \\
 Stability and excitations of the quantum paramagnetic phase
}

\author{R. L. Doretto}
\affiliation{Instituto de F\'isica Te\'orica,
             Universidade Estadual Paulista,
             01140-070 S\~ao Paulo, SP, Brazil}
\author{Matthias Vojta}
\affiliation{Institut f\"ur Theoretische Physik,
             Technische Universit\"at Dresden,
             01062 Dresden, Germany}

\date{\today}

\begin{abstract}
Motivated by experiments on non-magnetic triangular-lattice Mott
insulators, we study one candidate paramagnetic phase, the columnar
dimer (or valence-bond) phase. We apply variants of the bond-operator
theory to a dimerized and spatially anisotropic spin-1/2 Heisenberg
model and determine its zero-temperature phase diagram and the
spectrum of elementary triplet excitations (triplons).
Depending on model parameters, we find that the minimum of the triplon
energy is located at either a commensurate or an incommensurate
wavevector. Condensation of triplons at this
commensurate--incommensurate transition defines a quantum Lifshitz
point, with effective dimensional reduction which possibly leads to
non-trivial paramagnetic (e.g. spin-liquid) states near the closing of
the triplet gap.
We also discuss the two-particle decay of high-energy triplons, and we
comment on the relevance of our results for the organic Mott insulator
\pdmit.
\end{abstract}
\pacs{75.10.Jm, 75.10.Kt, 75.30.Kz, 75.50.Ee}

\maketitle

%
%
%

\section{Introduction}
\label{sec:intro}

In the search for novel phases, frustrated quantum antiferromagnets
have attracted enormous attention. Here, the combined effect of
geometric frustration and quantum fluctuations tends to destabilize
conventional magnetic order in favor of quantum paramagnetic ground
states, such as valence-bond solids (VBS) with broken lattice
symmetries or featureless spin liquids.\cite{review-subir,review-balents,misguich}

A prominent example of a frustrated quantum magnet is the spin-1/2
Heisenberg model on the triangular lattice. Although it was initially
proposed by Anderson\cite{rvb-triangle} that its ground state could be
a resonating-valence-bond (RVB) spin liquid, it is now
established that the model displays semiclassical non-collinear 120$^\circ$
order.\cite{isotropic-model}
However, modifications beyond the simple nearest-neighbor exchange are
believed to induce non-magnetic phases on the triangular lattice. For
instance, it has been argued\cite{schmidt10} that a combination of
longer-range and multiple-spin-exchange
interactions is important near the Mott transition to a metallic
state, where they induce a spin-liquid state as observed in numerical
studies of the triangular-lattice Hubbard model. Models with spatially
anisotropic exchange interactions have been studied as
well,\cite{trumper99,merino99,series-exp,ccm,rg,kallin11,spin-liquid,frg,xu09,weichsel11,hauke11}
interpolating from the triangular lattice to both the square-lattice
and the decoupled-chain limits, and both
VBS and spin-liquid phases have been proposed to occur.

Experimentally, triangular-lattice Heisenberg models are realized in a
variety of materials, such as the inorganic Mott
insulators Cs$_2$CuCl$_4$ and Cs$_2$CuBr$_4$ (Ref.~\onlinecite{cscucl,cscubr})
and the organic compounds $\kappa$-(ET)$_2$Cu$_2$(CN)$_3$
(Ref.~\onlinecite{kappa-et}) and \xdmit\
(Refs.~\onlinecite{dmit01,dmit02,dmit03,dmit04,dmit_rev}, with X=EtMe$_3$Sb or
EtMe$_3$P), all showing some degree of spatial anisotropy.
In the latter two, the crystal lattice is rather soft: pressure can be
used to induce an insulator-to-metal transition, with
superconductivity appearing at low temperature. Spin--lattice
interactions are also relevant in the insulator, as they potentially
relieve frustration via dimerization. In fact, such a dimerized VBS
phase has been reported for \pdmit, likely stabilized by
magneto-elastic coupling.

\begin{figure*}[!t]
\centerline{\includegraphics[width=10.0cm]{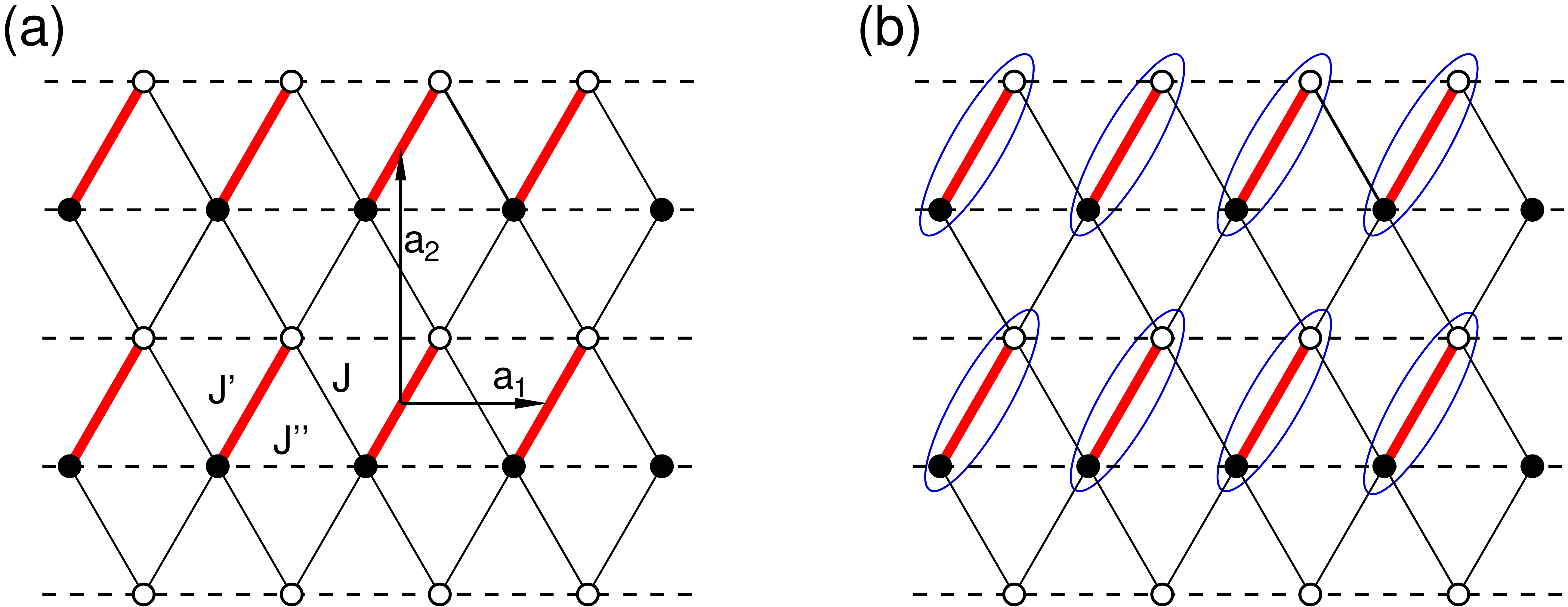}
            \hskip1.2cm
            \includegraphics[width=4.6cm]{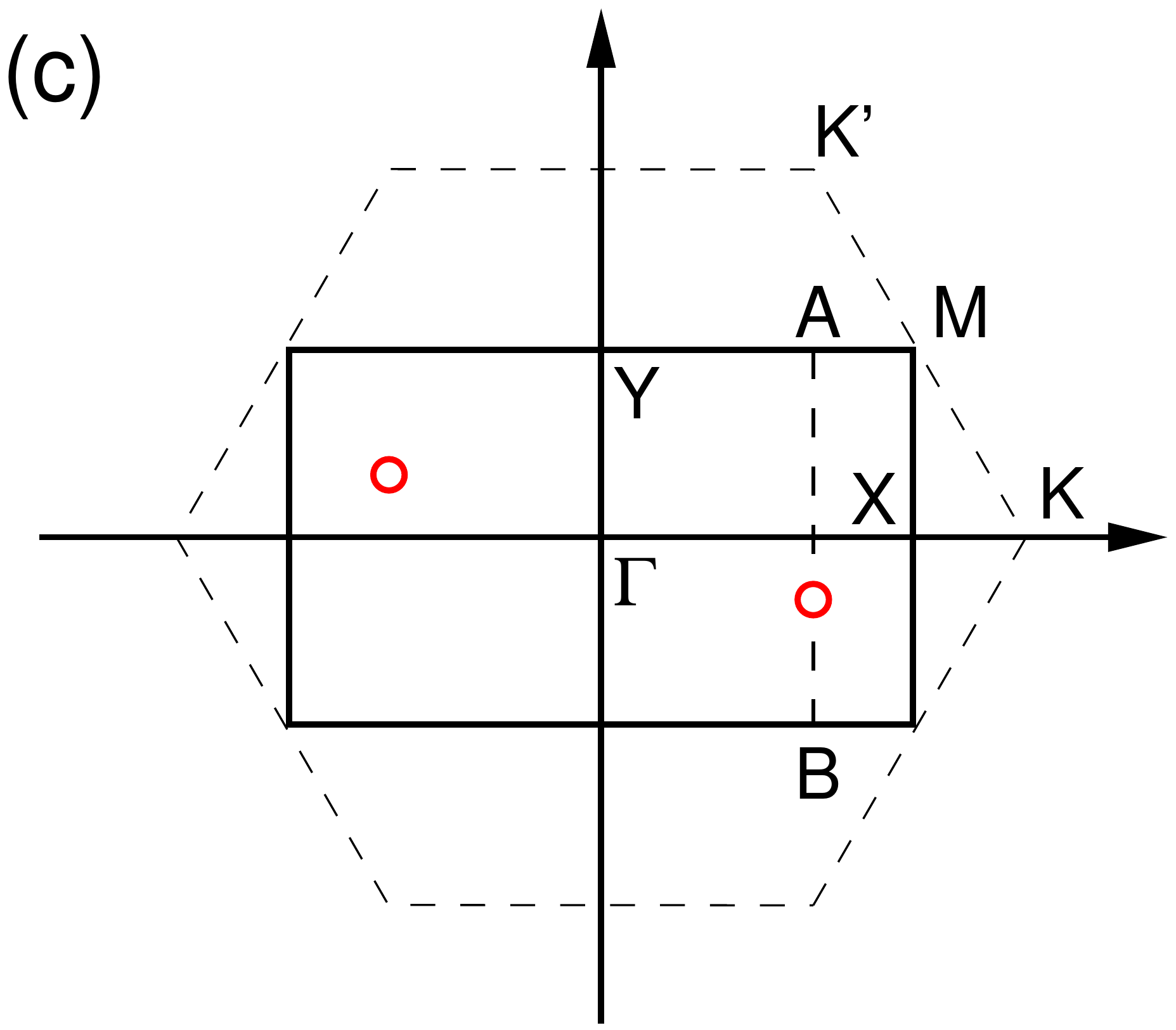}
}
\caption{(Color online)
 (a) Lattice structure of the model \eqref{ham}, with exchange couplings
 $J$, $J'$, and $J''$ shown as black (thin), red (thick), and dashed
 black lines, respectively. Solid (open) circles represent spin $\bS^1$
 ($\bS^2$) of each dimer. ${\bf a}_1$ and ${\bf a}_2$ are the
 primitive vectors of the dimerized lattice.
 (b) Dimerized ground state in the limit of large $J'$. Here, ellipses
 represents a singlet.
 (c) Brillouin zone of the dimerized lattice (solid), together with the
 (hexagonal) Brillouin zone of the original triangular lattice
 (dashed). The open circles indicate the minimum position of
 $\omega_\bq$ [Eq.~\eqref{omega-harmonic}] for parameters $J' = 1.5$
 and $J'' = 1$. Here ${\bf K} = (4\pi/3a,0)$, ${\bf K'} =
 (2\pi/3a,2\pi/\sqrt{3}a)$, ${\bf X} = (\pi/a,0)$,
 ${\bf Y} = (0,\pi/\sqrt{3}a)$, and  ${\bf M} = (\pi/a,\pi/\sqrt{3}a)$
 with $a$ being the lattice spacing of the triangular lattice.
}
\label{fig:model}
\end{figure*}

In this paper, we study the columnar dimer phase of a generic
triangular-lattice antiferromagnet (AF).\cite{pattern} In the absence of
detailed knowledge about microscopics, we focus on a simple, yet
very rich, model (Fig.~\ref{fig:model}a) with
anisotropic nearest-neighbor couplings and explicit dimerization, the
latter possibly arising from magneto-elastic effects.
A determination of the full phase diagram of this model (see
Fig.~\ref{fig:schem} for a sketch) is a hard task. Here we stick to
the more modest goal of characterizing the dimer phase and its triplon
excitations and locating the line of quantum phase transitions
(QPTs) to a magnetically ordered state, where the triplon gap
closes. To this end, we employ the bond-operator technique introduced
by Sachdev and Bhatt.\cite{SachdevBhatt} It turns out to be important
to go beyond the linearized (i.e. non-interacting) boson problem: We
discuss various approximation schemes, also providing a guideline as to
which non-linear effects are important in the presence of frustration.

Among our results is the existence of a commensurate--incommensurate
transition (CIT) in the excitation spectrum, i.e., the minimum of the
triplon dispersion is locked at wavevector $(0,0)$ (in the coordinates
of the dimerized lattice, Fig.~\ref{fig:model}) in some region of
parameter space, while it moves to incommensurate momenta elsewhere. If
the triplon gap closes at the CIT line, the excitations become
anomalously soft, leading to a quantum Lifshitz
point.\cite{qlif1,qlif2} We briefly discuss the unusual quantum
critical regime of such a putative magnetic QPT, but also speculate
about an intervening non-trivial paramagnetic phase.
Finally, we discuss two-particle (as opposed to three-particle)
decay of magnetic excitations,\cite{kole06,zhito06} which is
generically possible in low-symmetry dimerized magnets.

\subsection{Model}
\label{sec:model}

We consider a triangular-lattice Heisenberg model,
\begin{equation}
 \mathcal{H} = \sum_{\langle i, j \rangle}J_{ij}\bS_i\cdot\bS_j,
\label{ham}
\end{equation}
where $\bS_i$ represents a spin--1/2 at site $i$, and the
nearest-neighbor exchange couplings $J_{ij}$ equal $J$, $J'$,
and $J'' $ according to the pattern shown in
Fig.~\ref{fig:model}a.\cite{pattern} Hereafter we employ $J = 1$.

\subsection{Phase diagram}

\begin{figure}[!t]
\centerline{\includegraphics[width=8.6cm]{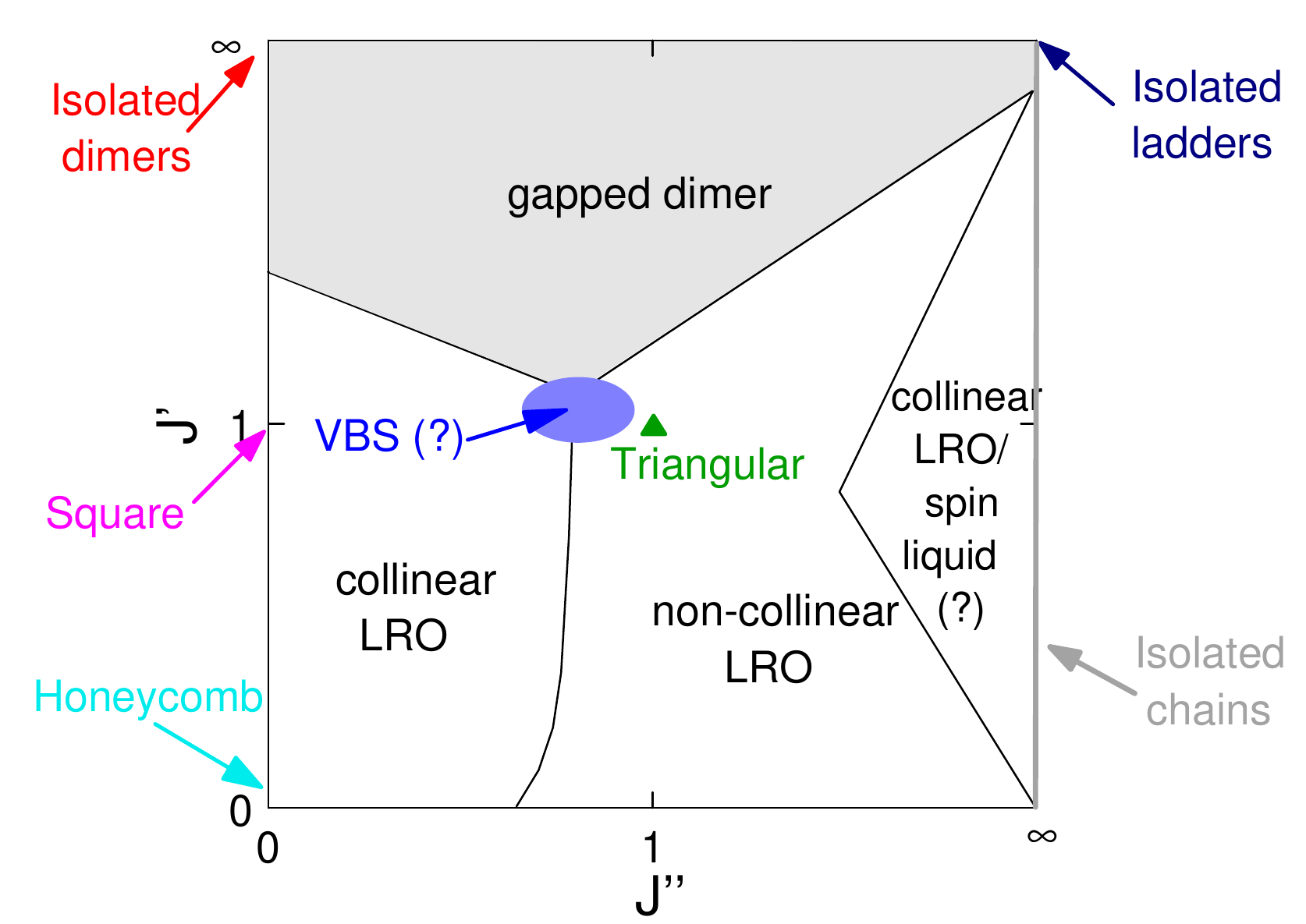}}
\caption{(Color online)
  Schematic phase diagram of the Heisenberg model \eqref{ham}, as
  function of the couplings $J'$ and $J''$, keeping $J=1$. For large
  $J'$, a gapped dimer phase is realized (shaded). Various other
  limits are indicated in the figure, see text for details.
  The multicritical point, where the collinear LRO, non-collinear LRO,
  and gapped dimer phases meet, is the quantum Lifshitz point.
  [The classical limit of \eqref{ham} is discussed in Appendix~\ref{ap:class}.]
}
\label{fig:schem}
\end{figure}

To appreciate the richness of the model \eqref{ham}, it is useful to
discuss a few limiting cases, with their properties summarized in the
schematic phase diagram in Fig.~\ref{fig:schem}.

(i) Line $J'' = 1$. On this line, for $J' = 1$, we recover the isotropic
triangular-lattice AF Heisenberg model. Its ground state has
non-collinear $120^\circ$ N\'eel order, with {\it commensurate}
ordering wavevectors ${\bf K} = (4\pi/3a,0)$ and
${\bf K'} = (2\pi/3a,2\pi/\sqrt{3}a)$.\cite{isotropic-model}
Non-collinear order can be expected to extend to small values of
$J'$. On the other hand, the limit $J' \gg 1$ corresponds to a
paramagnet of weakly coupled dimers,\cite{gia08} such that spins which
are pairwise coupled by $J'$ combine into a columnar arrangement of
singlets, as illustrated in Fig.~\ref{fig:model}b.

(ii) Line $J'' = 0$. Model \eqref{ham} is now topologically equivalent
to the staggered dimerized Heisenberg model on a square lattice,
recently discussed in
Refs.~\onlinecite{wenzel08,stag-dimer}. According to quantum Monte
Carlo (QMC) calculations, a order-disorder QPT takes place at $J'_{c}
= 2.5196$,\cite{wenzel08} with collinear N\'eel order setting in for
$J' < J'_{c}$. The line $J''=0$ also includes the
square and honeycomb lattice AF at $J'=1$ and 0, respectively.

(iii) Line $J' = 1$. Here, Eq.~\eqref{ham} is nothing else but the
spatially anisotropic triangular-lattice AF addressed in
Refs.~\onlinecite{trumper99,merino99,series-exp,ccm,rg,spin-liquid,frg,xu09,kallin11,weichsel11,hauke11}
In particular, coupled-cluster calculations\cite{ccm} indicate that
the system displays collinear N\'eel order for $J'' < 0.8$ and
non-collinear spiral order with an {\it incommensurate} ordering
wavevector (except at $J'' = 1$) for $0.8 < J'' < 1.8$. For
$J''\gg 1$ the system consists of weakly coupled chains, and a collinear
AF state\cite{ccm} has been argued to arise as a result of
order-from-disorder physics.\cite{rg,kallin11} However, we note that both
non-collinear\cite{weichsel11} and disordered (i.e. spin-liquid)
ground states\cite{spin-liquid,frg} have also been
proposed in this regime. For $0.7<J''<0.9$ the physics is under debate
as well: series-expansion studies favored a spontaneously dimerized
VBS in this region,\cite{series-exp} which was not found in the
coupled-cluster study.\cite{ccm}

In total, the phase diagram of the model \eqref{ham},
Fig.~\ref{fig:schem}, displays a gapped paramagnetic phase, both
collinear and non-collinear long-range order (LRO), as well as
putative non-trivial spin-liquid
regimes which may or may not be adiabatically connected
to the one-dimensional limit $J''\to\infty$.

In this paper, our focus is on the properties of the dimerized
paramagnet which is adiabatically connected to the limit of large
$J'$. As stated above, in a real system, such as \pdmit\ the
dimerization may arise spontaneously due to longer-range or
ring-exchange couplings, and will then be stabilized by
magneto-elastic effects. Leaving a detailed study of the latter for
the future, we choose to work with explicit
dimerization as in Fig.~\ref{fig:model}.\cite{pattern}
Most of our calculations are restricted to the parameter range $J' >
J''$ and $0 \le J'' \le 1.5$. A quantitative phase diagram
obtained  from various bond-operator approximations is in
Fig.~\ref{fig:phase-diag} and will be discussed in detail below.

\subsection{Outline}

\begin{figure}[!t]
\centerline{\includegraphics[width=8.2cm]{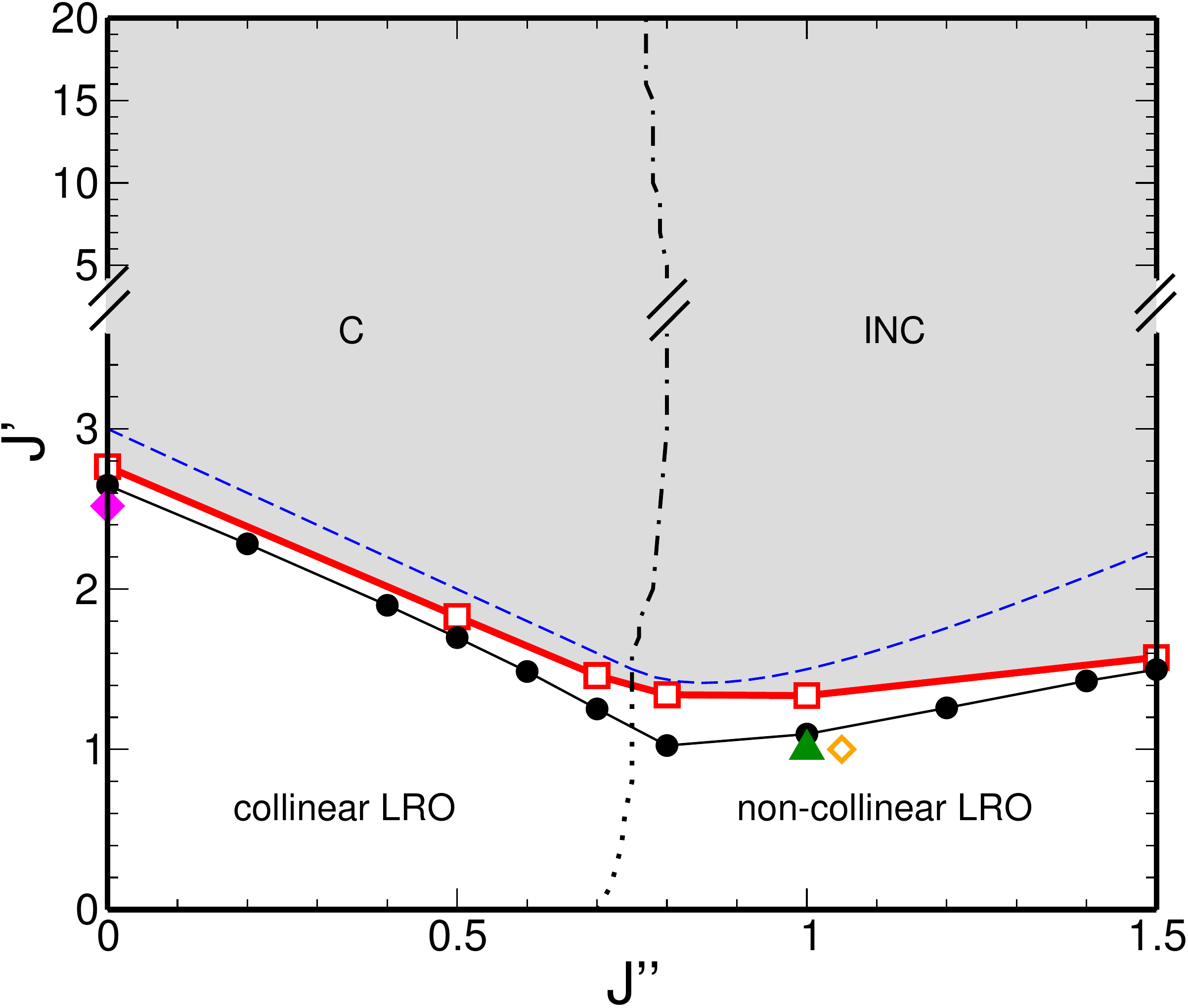}}
\caption{(Color online)
Quantitative phase diagram for the Heisenberg model \eqref{ham}, obtained using the
bond-operator formalism. The dimer phase boundary (${\color{red}\square}$) has been
obtained from the HF-cubic approximation described in the paper; for comparison we also
show the results of the harmonic mean-field\cite{SachdevBhatt}
($\CIRCLE$) and harmonic spin-wave approximations (dashed blue).\cite{kotov98}
The dashed-dotted line separates the commensurate (C) and incommensurate (INC) regions of
the dimer phase at large $J'$. It continues into the dotted line which separates the
phases with collinear and non-collinear LRO at small $J'$ -- note that this line is an
estimate based on series-expansion\cite{series-exp} and classical-limit results
(Appendix~\ref{ap:class}.)
Also shown are the isotropic point $J' = J'' = 1$ (${\color{darkgreen}\blacktriangle}$) and the
point $J' = 1$, $J'' = 1.05$ (${\color{orange}\bf\lozenge}$) for \pdmit.\cite{dmit02} (Here,
longer-range or multiple-spin interactions can be expected to be relevant in addition.)
Finally, on the $J''=0$ axis we have indicated the QMC value of $J'_c$
(${\color{magenta}\blacklozenge}$) for the square-lattice staggered dimerized Heisenberg
model.\cite{wenzel08}
}
\label{fig:phase-diag}
\end{figure}

The remainder of our paper is organized as follows:
In Sec.~\ref{sec:effective-model}, we briefly summarize the
bond-operator formalism that is employed to describe the paramagnetic
phase of the model \eqref{ham} and derive an effective
Hamiltonian of interacting triplet excitations (triplons).
In Sec.~\ref{sec:harmonic}, we analyze the non-interacting
(i.e. harmonic) part of this Hamiltonian. We determine the magnitude
and momentum-space location of the minimum energy
gap of the triplons.
Corrections arising from interactions -- both cubic and quartic --
among the triplons are analyzed in some detail in
Sec.~\ref{sec:interaction}.  For all levels of approximation,
the closing of the triplon gap allows to construct the phase boundary
between the singlet and long-range-ordered phases, shown in
Fig.~\ref{fig:phase-diag}.
In Sec.~\ref{sec:discussion}, we discuss various aspects of our
results, such as the two-particle decay of triplons due to cubic
interactions, the commensurate--incommensurate
transition, and the associated quantum Lifshitz point. We also comment
on some implications for the organic Mott insulator \pdmit.
Concluding remarks close the paper. A summary of the classical phase
diagram of \eqref{ham} as well as details of the calculations are
relegated to the appendices.

\section{Bond operators and triplon excitations}
\label{sec:effective-model}

For $J'>J,J''$ it is useful to view the spin pairs coupled by $J'$
as building blocks of the model. The four states per such dimer can be
efficiently represented using bond operators.\cite{SachdevBhatt} which
naturally lead to a description of the elementary excitations of the
paramagnetic phase in terms of bosonic spin-1 modes, dubbed
``triplons''.

\subsection{Bond-operator representation}
\label{sec:bondop}

To introduce bond operators, the triangular lattice of spins is
re-interpreted as rectangular lattice of dimers,
Fig.~\ref{fig:model}b, with sites $i$.
For each dimer, one introduces bosonic operators
$\{s_i^\dagger,t^\dagger_{i\alpha}\}$ ($\alpha=x,y,z$), which create the dimer states out
of a fictitious vacuum. Explicitly (and omitting the site index $i$), $|s\rangle =
s^\dagger |0\rangle$, $|\alpha\rangle = t^\dagger_{\alpha}|0\rangle$, where
$\left|s\rangle\right. =
(\left|\uparrow\downarrow\rangle\right.-\left|\downarrow\uparrow\rangle\right.)/\sqrt{2}$,
$\left|x\rangle\right. =
(-\left|\uparrow\uparrow\rangle\right.+\left|\downarrow\downarrow\rangle\right.)/\sqrt{2}$,
$\left|y\rangle\right. =
i(\left|\uparrow\uparrow\rangle\right.+\left|\downarrow\downarrow\rangle\right.)/\sqrt{2}$,
$\left|z\rangle\right. =
(\left|\uparrow\downarrow\rangle\right.+\left|\downarrow\uparrow\rangle\right.)/\sqrt{2}$.
The Hilbert-space dimension is conserved by imposing the constraint
\begin{equation}
\label{constraint}
s_i^\dagger s_i +
\sum_\alpha  t^\dagger_{i\alpha} t_{i\alpha} = 1
\end{equation}
on every site $i$. The original spin operators $\bS^1$ and $\bS^2$ of each dimer are
given by
\begin{eqnarray}
S^{1,2}_\alpha &=& \pm\frac{1}{2}\left(s^\dagger t^{\phantom{\dagger}}_\alpha
              + t^\dagger_\alpha s
          \mp i\epsilon_{\alpha\beta\gamma}t^\dagger_\beta t^{\phantom{\dagger}}_\gamma \right),
\label{spin-bondop}
\end{eqnarray}
where $\epsilon_{\alpha\beta\gamma}$ is the antisymmetric
tensor with $\epsilon_{xyz} = 1$ and summation convention over
repeated indices is implied.

\subsection{Effective theory for triplet excitations}
\label{sec:model-bond}

Expressing the Hamiltonian \eqref{ham} in terms of the dimer spins
($\bS^1_i$ and $\bS^2_i$) yields
\begin{eqnarray}
\mathcal{H} &=& \sum_i \left[ J'\bS^1_i\cdot\bS^2_i  +
      J''\left(\bS^1_i\cdot\bS^1_{i+1} +  \bS^2_i\cdot\bS^2_{i+1}\right)
      \right]
\nonumber \\
             &+& J \sum_{i,n}\bS^2_i\cdot\bS^1_{i+n}.
\label{ham-underline}
\end{eqnarray}
Here $n = 1,2,3$ corresponds to the nearest-neighbor vectors
\begin{equation}
 \taub_1 = a\hat{x}, \;\;\;\;\;\;\;
 \taub_2 = \sqrt{3}a\hat{y}, \;\;\;\;\;\;\;
 \taub_3 = a(\hat{x} + \sqrt{3}\hat{y}),
\label{tauvectors}
\end{equation}
with $a$ being the lattice spacing of the {\it original} triangular
lattice (in the following $a=1$).
Substituting Eq.~\eqref{spin-bondop} into \eqref{ham-underline}
yields a Hamiltonian of the form
\begin{equation}
  \mathcal{H} = \mathcal{H}_0 + \mathcal{H}_2 + \mathcal{H}_3 + \mathcal{H}_4
\label{ham-bond}
\end{equation}
where $\mathcal{H}_n$ contains $n$ triplet operators (see Appendix \ref{ap:details1} for
details).

In the paramagnetic phase realized for large $J'$, the singlet can be viewed as
``condensed'',\cite{SachdevBhatt} formally $ s^\dagger_i, s_i \rightarrow \sqrt{N_0}$ in
Eq.~\eqref{ham-bond}. As a consequence, one ends up with an effective Hamiltonian solely
in terms of the (bosonic) triplet operators $t^\dagger_{i\,\alpha}$. In the spirit of
mean-field theory, the constraint \eqref{constraint} is implemented on average via a
Lagrange multiplier $\mu$. Both $\mu$ and $N_0$ will be self-consistently
determined. One expects that $N_0 \approx 1$ in the limit $J' \gg J,J''$.
(Other bond-operator schemes will be discussed in Sec.~\ref{sec:bondop2} below.)

After performing a Fourier transform, i.e.,
$t^\dagger_{i\,\alpha} = N'^{-1/2}\sum_\bk \exp(-i\bk\cdot\bR_i)t^\dagger_{\bk\alpha}$,
the terms of the Hamiltonian \eqref{ham-bond} read
\begin{eqnarray}
E_0 &=& -3J'N/8 - \mu N(N_0 - 1)/2,
\nonumber \\
&& \nonumber \\
\mathcal{H}_2 &=& \sum_\bk \left[ A_\bk t^\dagger_{\bk\alpha}t_{\bk\alpha}
                  + \frac{1}{2}B_\bk \left( t^\dagger_{\bk\alpha}t^\dagger_{\bk\alpha}
                  + {\rm H.c.}\right) \right],
\label{hharmonic2} \\
&& \nonumber \\
\mathcal{H}_3 &=& \frac{1}{2\sqrt{N'}}\epsilon_{\alpha\beta\lambda}\sum_{\bp,\bk}\xi_{\bk-\bp}
                  \; t^\dagger_{\bk-\bp\alpha}t^\dagger_{\bp\beta}t_{\bk\lambda} + {\rm H.c.},
\label{hcubic2} \\
&& \nonumber \\
\mathcal{H}_4 &=& \frac{1}{2N'}\epsilon_{\alpha\beta\lambda}\epsilon_{\alpha\mu\nu}
                  \sum_{\bq,\bp,\bk} \gamma_\bk \;
                   t^\dagger_{\bp+\bk\beta}t^\dagger_{\bq-\bk\mu}t_{\bq\nu}t_{\bp\lambda}.
\label{hquartic2}
\end{eqnarray}
Here $N' = N/2$, with $N$ being the number of sites of the original
triangular lattice, and the momentum sum runs over the dimerized
(rectangular) Brillouin zone. The coefficients $A_\bk$, $B_\bk$, $\xi_\bk$,
and $\gamma_\bk$ are given by
\begin{eqnarray}
 A_\bk      &=& \frac{J'}{4} - \mu + B_\bk,
\label{coef-a}\\
&& \nonumber \\
 B_\bk      &=& \frac{1}{2}N_0\left[(2J'' - 1)\cos k_x - \cos(\sqrt{3}k_y) \right.
\nonumber \\
            && \left. \;\;\;\;\;\;\;\;\;\;\;  - \cos(k_x + \sqrt{3}k_y)\right],
\label{coef-b}\\
&& \nonumber \\
 \xi_\bk    &=& -\sqrt{N_0}\left[\sin k_x + \sin(\sqrt{3}k_y) +
                \sin(k_x + \sqrt{3}k_y)\right],
\nonumber \\
&& \label{xi} \\
 \gamma_\bk &=& -\frac{1}{2}\left[(2J'' + 1)\cos k_x   + \cos(\sqrt{3}k_y) \right.
\nonumber \\
            && \left. \;\;\;\;\;\;\;\;\;\;\; + \cos(k_x + \sqrt{3}k_y)\right].
\label{gamma}
\end{eqnarray}

A few remarks about the general structure of the effective Hamiltonian \eqref{ham-bond}
are here in order: The bond-operator approach has some parallels to the
Holstein-Primakoff approach to ordered magnets, with the difference that one considers
fluctuations above a quantum paramagnetic state instead of a N\'eel state. A cubic
triplet term \eqref{hcubic2} is present in many low-symmetry situations, including the
model considered here and also the square-lattice staggered dimerized Heisenberg model of
Refs.~\onlinecite{wenzel08,stag-dimer}. This is to be contrasted with spin waves where a
cubic interaction term -- describing two-particle decay of transverse magnons -- is only
present for non-collinear spin structures while it vanishes for collinear ones (see
Ref.~\onlinecite{cherny09} for a discussion).

\begin{figure*}[t]
\centerline{\includegraphics[height=6.1cm]{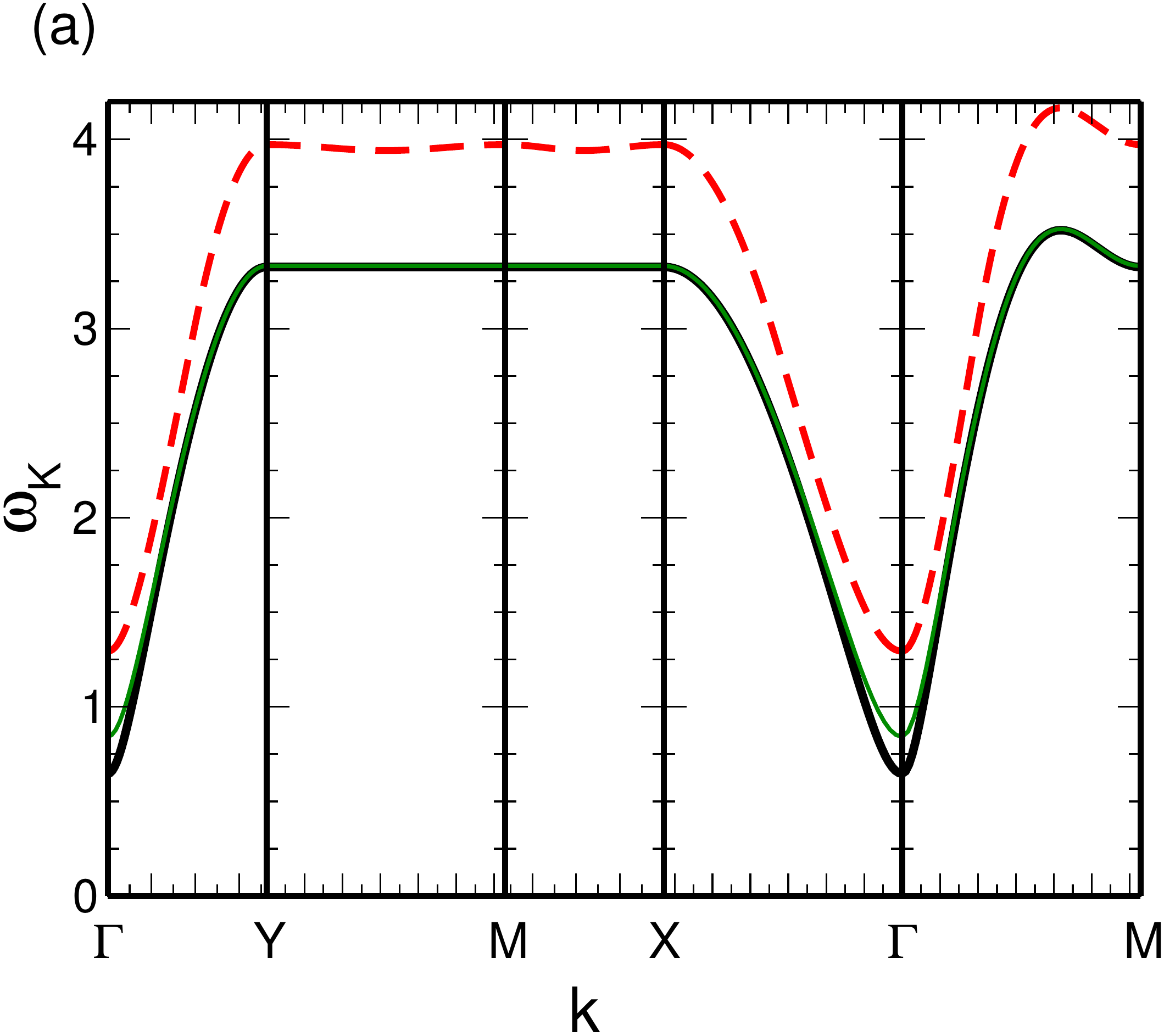}
            \hskip0.2cm
            \includegraphics[height=6.1cm]{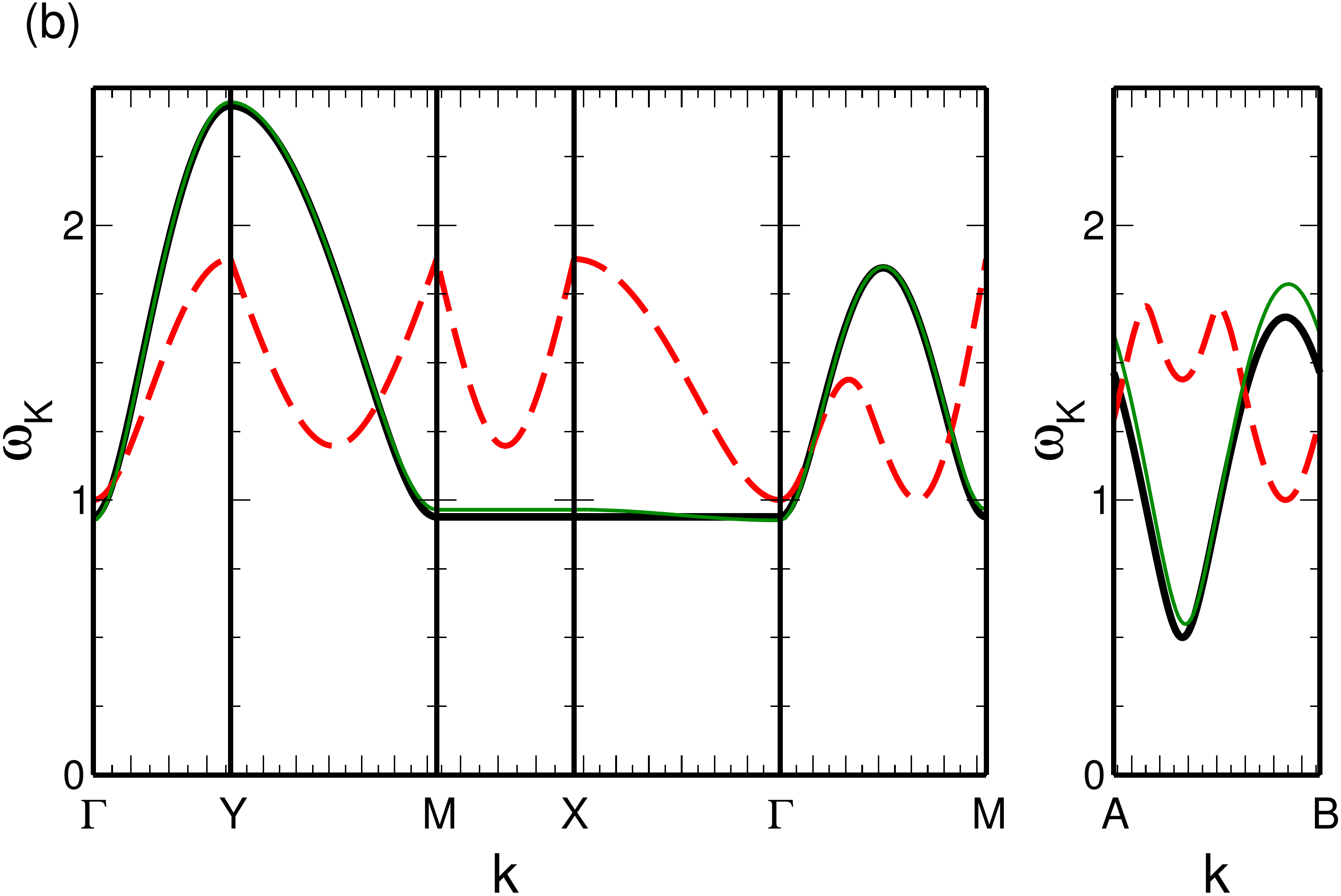}}
\vskip-0.2cm
\centerline{\includegraphics[width=8.1cm]{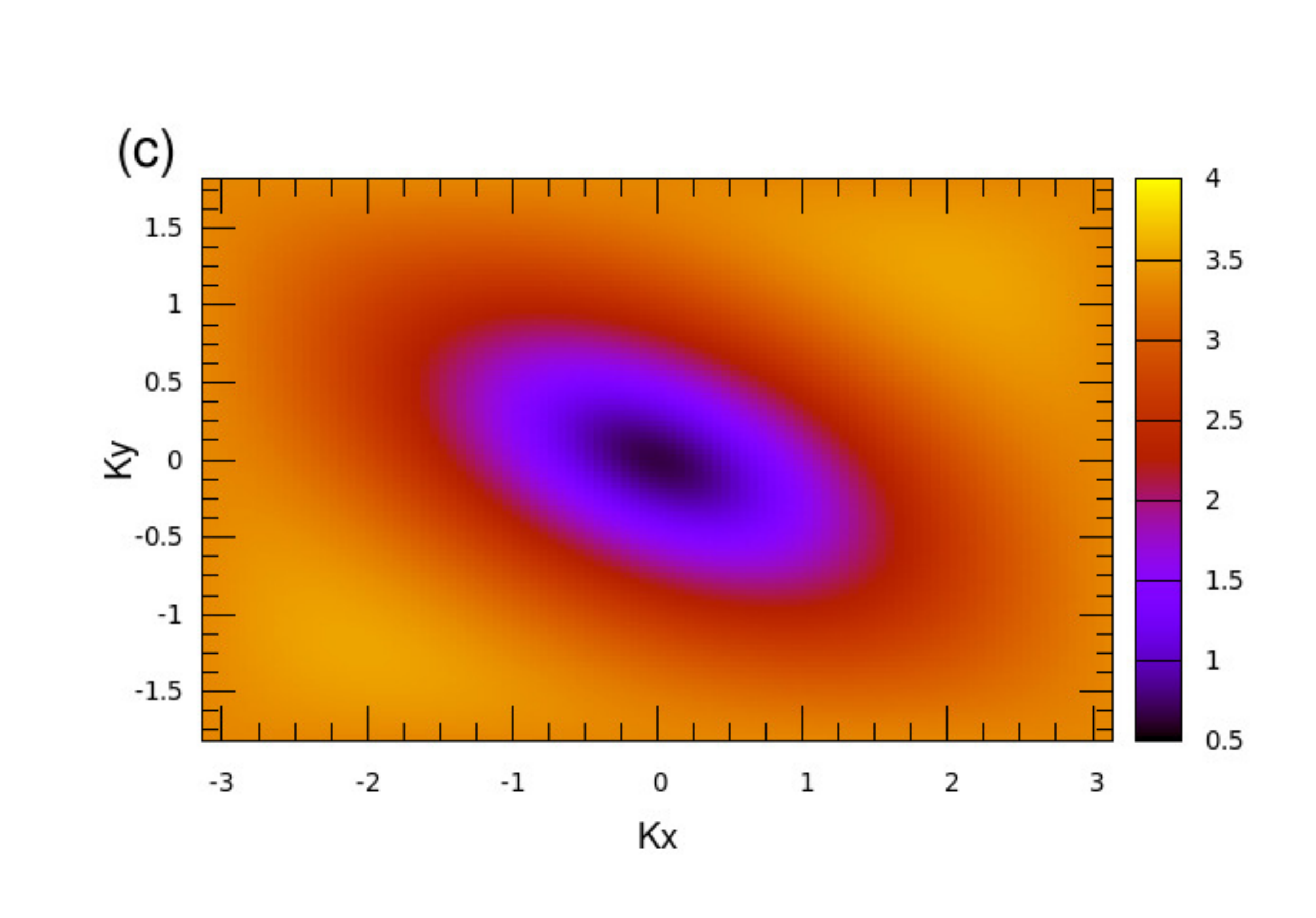}
            \hskip-0.1cm
            \includegraphics[width=8.1cm]{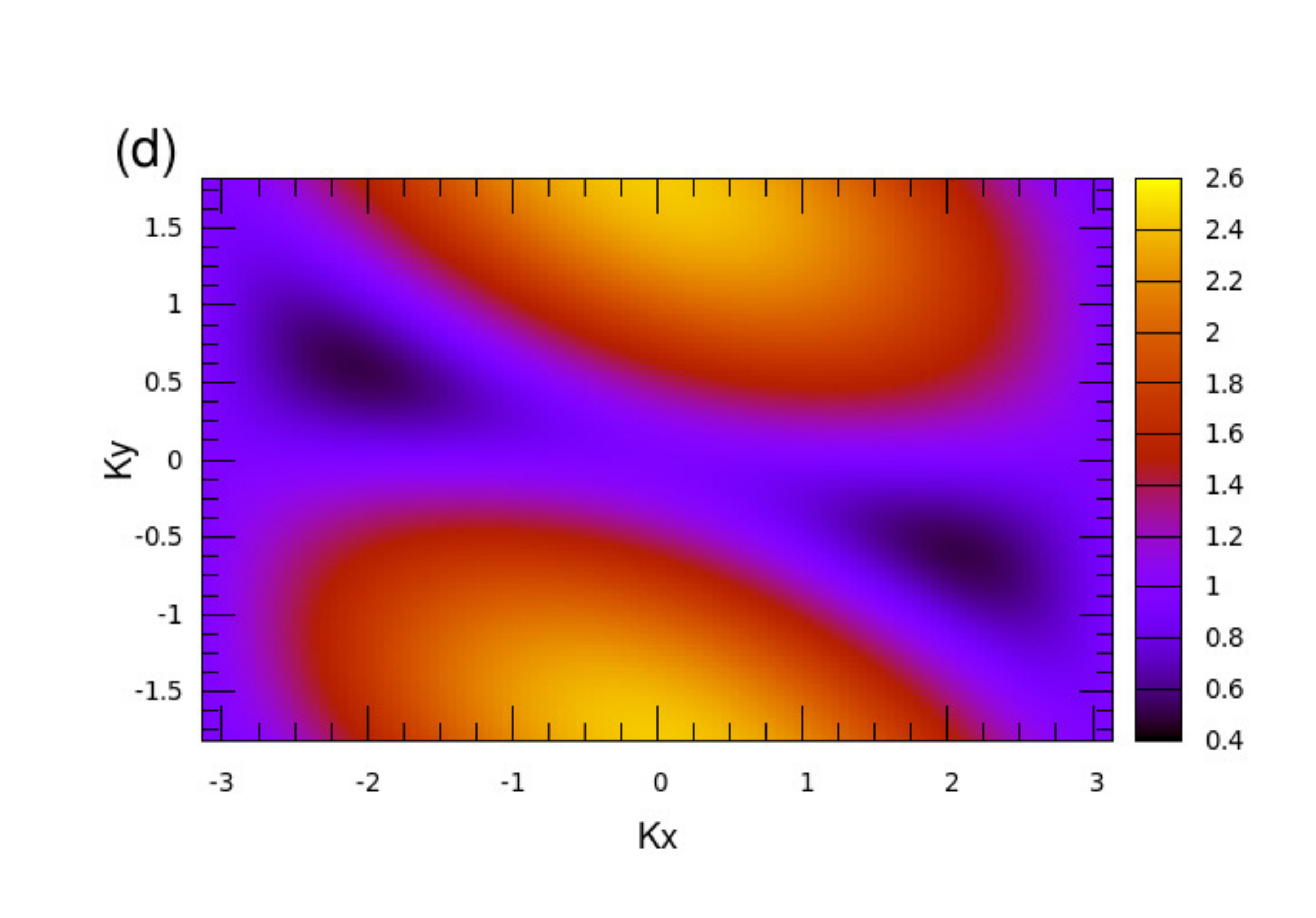}}
\vskip-0.3cm

\caption{(Color online)
 Upper row: Triplon dispersion relation at the harmonic level (solid
 black) along paths in the dimerized Brillouin zone
 (Fig.~\ref{fig:model}~c), for (a) $J' = 3.0$, $J'' = 0$
 and (b) $J' = 1.5$, $J'' = 1$. For comparison, the dispersion
 obtained within the HF approximation for the quartic term (thin
 green line, see Sec.~\ref{sec:quartic}) is also shown, indicating
 that the corrections from the quartic term are minor. Finally, the
 dashed red line represents the bottom
 of the two-particle continuum at the harmonic level.
 Lower row: Contour plot of the triplon energy at the harmonic level,
 for (c) $J' = 3.0$, $J'' = 0$ and (d) $J' = 1.5$, $J'' = 1$.
}
\label{fig:disp-harmonic}
\end{figure*}

\subsection{Alternative bond-operator schemes}
\label{sec:bondop2}

The procedure discussed in the previous section, where the constraint \eqref{constraint}
is included into the description via a Lagrange multiplier and $N_0$ is self-consistently
determined, is the one originally proposed by Sachdev and Bhatt.\cite{SachdevBhatt}
However, alternative methods to deal with the constraint \eqref{constraint} can be found
in the literature. Let us briefly summarize two of them.

Chubukov and Morr\cite{chubukov95} resolve the constraint via $s =
(1 - \lambda t^\dagger_\alpha t_\alpha)^{1/2}$ where $\lambda$ is an artificial control
parameter, with $\lambda=1$ corresponding to the physical case. The square root can now
be expanded to obtain a Taylor series in the control parameter
$\lambda$ -- a procedure similar to the $1/S$ expansion in
conventional spin-wave theory. This generates a Hamiltonian
with triplon terms up to arbitrary order, which can be analyzed order by order in
$\lambda$. Finally, one arrives at physical results by taking the limit $\lambda \to 1$.

Kotov {\it et al.}\cite{kotov98} instead implement the hard-core constraint explicitly in
the following way. First, the $t$ operators are re-interpreted as creation operators of
triplons on top of a singlet background. This is formally equivalent to setting
$s_i^\dagger=s_i=1$. The resulting Hamiltonian contains terms up to quartic order; its
harmonic piece, $\mathcal{H}_2$, is analogous to linear spin-wave theory.
The constraint is converted into the inequality
$\sum_\alpha  t^\dagger_{i\alpha} t_{i\alpha} \leq 1$, which
can be imposed using an infinite on-site triplet repulsion term
\[
\mathcal{H}_U = \frac{1}{2N'}U\sum_{i,\alpha,\beta}
                   t^\dagger_{i,\alpha}t^\dagger_{i,\beta}t_{i,\beta}t_{i,\alpha},
\;\;\;\;\;\;\;\;\;\;\;
U \rightarrow \infty.
\]
$\mathcal{H}_U$,
which provides the main contribution to the renormalization of the
noninteracting triplet energy, is treated using a self-consistent ladder summation of
diagrams (Brueckner approximation).

Both schemes were applied to the non-frustrated square-lattice bilayer Heisenberg model,
with the Brueckner approach\cite{kotov98} giving very accurate results, e.g., for
the location of the phase boundary. Results of similar quality where obtained for other
models with collinear spin correlations.\cite{kotov99}
In contrast, we have found that the Brueckner approach is less well suited for the
present triangular-lattice model: Following Ref.~\onlinecite{kotov98}, we have implemented
the Brueckner approximation for the hard-core triplon repulsion and included a
non-self-consistent treatment of the cubic term to second order. In the resulting phase
diagram, the stability of the paramagnetic phase is clearly overestimated, i.e., we find
the gapped paramagnet to be stable even at the isotropic point $J'=J''=1$, with the
critical $J'_c\approx0.75$ at $J''=1$.
We suspect that the combined effect of the hard-core and cubic terms in the presence of
non-collinear correlations requires a more accurate approximation, but have not explored
this route further.
Therefore we resort to the mean-field-based approach of
Ref.~\onlinecite{SachdevBhatt} whose results are presented in the body of the paper.

\begin{figure*}[t]
 \centerline{
  \includegraphics[width=7.1cm]{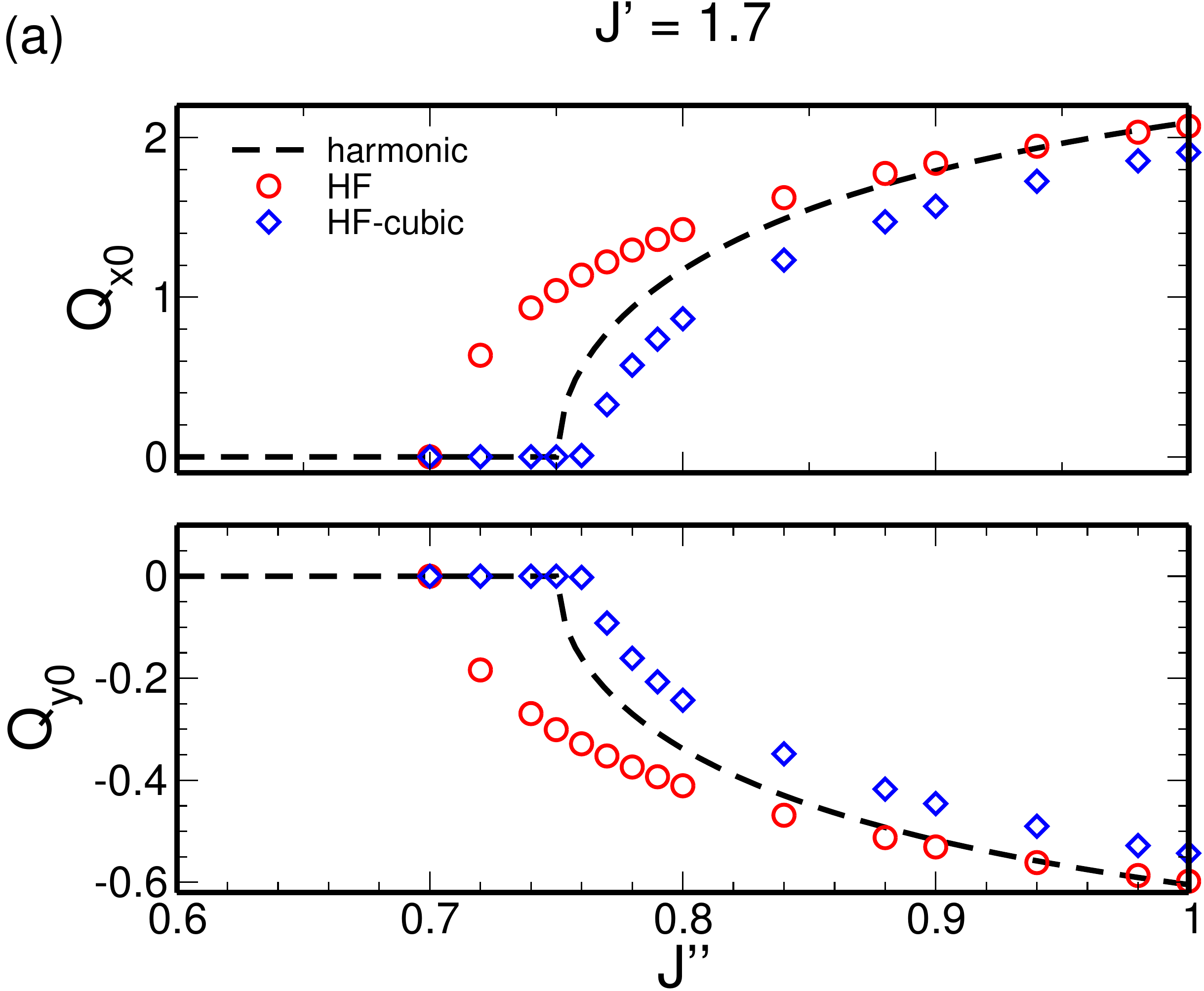}
  \hskip1.5cm
  \includegraphics[width=7.1cm]{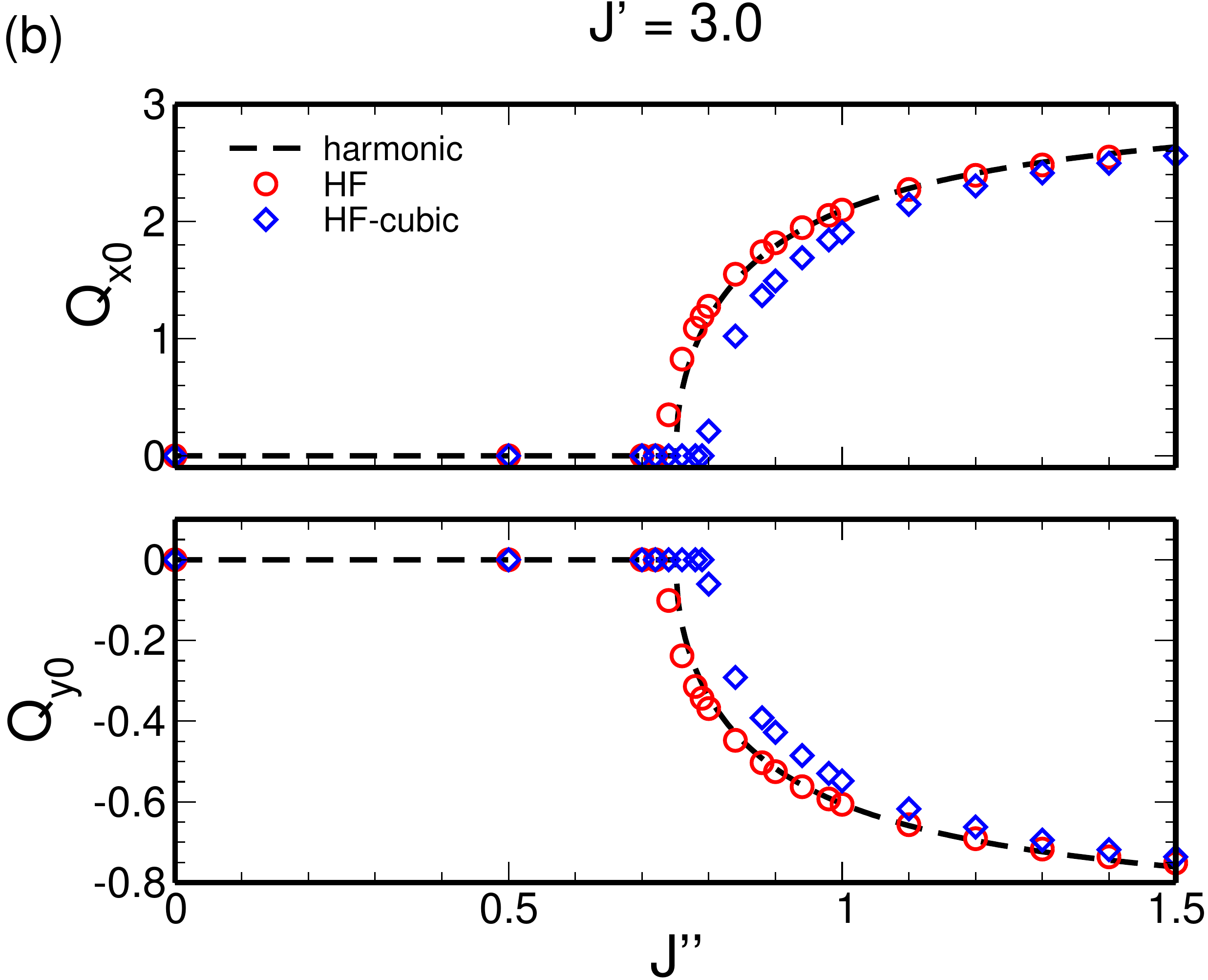}}
 \caption{(Color online) $x$ and $y$ components of the momentum $\bQ_0$
  (see text for definition) as a function of $J''$ at the harmonic (dashed),
  HF (${\color{red}\Circle}$), and HF-cubic (${\color{blue}\lozenge}$)
  approximations for the lines $J' = 1.7$ (a) and $3.0$ (b). Non-zero values of $\bQ_0$
  correspond to incommensurate correlations.
}
\label{fig:min01}
\end{figure*}

\begin{figure}[b]
 \includegraphics[width=7.1cm]{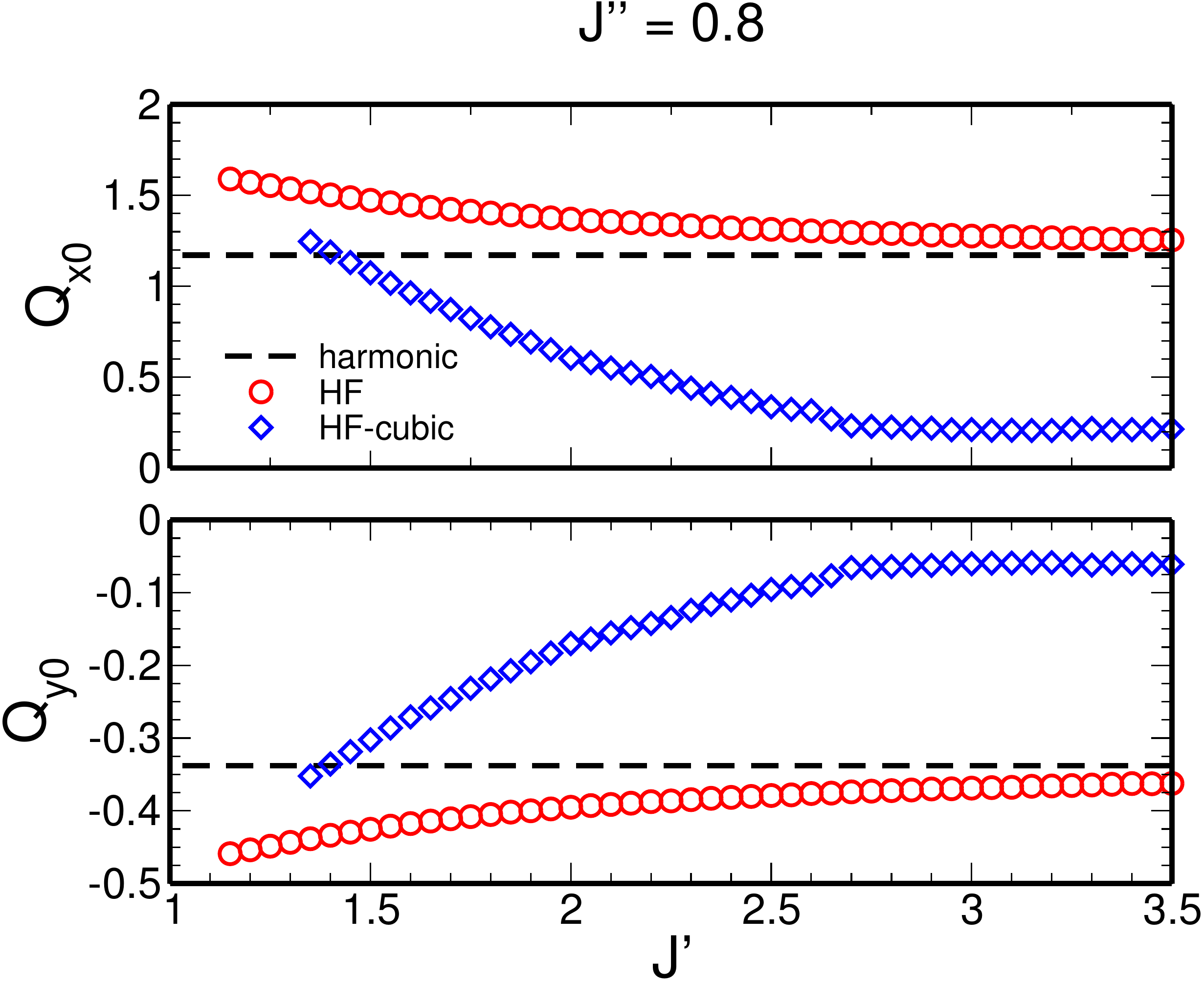}
 \caption{(Color online) $x$ and $y$ components of the momentum $\bQ_0$
  (see text for definition) as a function of $J'$ at the harmonic (dashed),
  HF (${\color{red}\Circle}$), and HF-cubic (${\color{blue}\lozenge}$)
  approximations for the line $J'' = 0.8$.
  In the strong-dimer limit of large $J'$, the harmonic $\bQ_0$
  is approached for both the HF and HF-cubic results.
}
\label{fig:min}
\end{figure}

\section{Harmonic approximation}
\label{sec:harmonic}

The lowest-order approximation to the triplon dynamics consists in keeping the quadratic
term $\mathcal{H}_2$ of the Hamiltonian only, which describes the physics of
noninteracting bosons. $\mathcal{H}_2$ can be diagonalized with
the help of the Bogoliubov transformation
\begin{equation}
  t^\dagger_{\bk\alpha} = u_\bk b^\dagger_{\bk\alpha} - v_\bk
                          b_{-\bk\alpha}.
\label{bogo-transf}
\end{equation}
One finds
\begin{equation}
\mathcal{H} = \bar{E}_0
            + \sum_{\bk\,\alpha} \omega_\bk
            b^\dagger_{\bk\alpha}b_{\bk\alpha},
\label{diag-ham}
\end{equation}
where
\begin{equation}
\bar{E}_0 = -\frac{3}{8}J'NN_0 - \frac{1}{2}\mu N(N_0 - 1)
     + \frac{3}{2}\sum_\bk \left(\omega_\bk - A_\bk \right)
\label{egs-harmonic}
\end{equation}
is the ground state energy,
\begin{equation}
 \omega_\bk = \sqrt{A^2_\bk - B^2_\bk}
\label{omega-harmonic}
\end{equation}
is the energy of the triplet excitations, and the Bogoliubov
coefficients in Eq.~\eqref{bogo-transf} obey
\begin{equation}
 u^2_\bk , v^2_\bk = \pm 1/2 + A_\bk/2\omega_\bk
 \;\;\;\; {\rm and} \;\;\;\;
 u_\bk v_\bk = B_\bk/2\omega_\bk.
\label{bogo-coef}
\end{equation}
From the saddle points conditions $\partial E_0/\partial N_0
= 0$ and $\partial E_0/\partial \mu = 0$, self-consistent equations
for $\mu$ and $N_0$ follow (see Appendix \ref{ap:details2} for
details)
\begin{eqnarray}
   \mu &=& -\frac{3J'}{4} + \frac{3}{N_0}\sum_\bk B_\bk v_\bk
           \left(v_\bk - u_\bk \right),
\label{self-mu} \\
&& \nonumber \\
   N_0 &=&  1 - \frac{3}{N'}\sum_\bk v^2_\bk =
            1 - \frac{1}{N'}\sum_{\bk\,\alpha} \langle
                  t^\dagger_{\bk\alpha}t_{\bk\alpha} \rangle.
\label{self-nzero}
\end{eqnarray}
Once $\mu$ and $N_0$ are known, the triplon energy \eqref{omega-harmonic}
is completely determined.


In Fig.~\ref{fig:disp-harmonic}, we show the triplon dispersion
relation for two sets of model parameters inside the disordered
phase. For the case $J' = 3$ and $J'' = 0$, the minimum gap is located
at ${\bf \Gamma}$, the center of the dimerized Brillouin zone,  while for
$J' = 1.5$ and $J'' = 1.0$, the minimum gap is at {\it incommensurate} momenta,
$\bQ_0 = \pm (2\pi/3,-\pi/3\sqrt{3})$.
Indeed, within the harmonic approximation, the momentum $\bQ_0$ associated
with the triplon gap can be analytically calculated. From
the solution of $\nabla_\bq\omega_\bq = \nabla_\bq B_\bq = 0$,
one finds that it depends only on the coupling $J''$, namely
\begin{equation}
\bQ_0 = \left\{ \begin{array}{cc}
          (0,0),  &  0 \le J'' \le 0.75, \\
          \pm ({q_{\rm xm}},{q_{\rm ym}}), &  J'' > 0.75,
                \end{array} \right.
\label{min-pos}
\end{equation}
with
\begin{eqnarray}
  {q_{\rm xm}} &=& 2\arccos\left(\frac{1}{4J'' - 2}\right),
\nonumber \\
  {q_{\rm ym}} &=& - \frac{1}{\sqrt{3}}\arccos\left(\frac{1}{4J'' -
  2}\right).
 \label{comp-qzero}
\end{eqnarray}
Note that $\bQ_0$ continuously moves from a commensurate, $(0,0)$, to an incommensurate
point, $\pm (q_{\rm xm},q_{\rm ym})$, as $J''$ increases, defining a
commensurate-incommensurate transition (CIT) in the excitation spectrum at $J''_{\rm
CIT} = 0.75$, see Fig.~\ref{fig:min01}.
At the present harmonic level, $\bQ_0$ only depends on $J''$, but this does not hold once
corrections are included, see Sec.~\ref{sec:interaction}. Further aspects of the CIT will
be discussed in Sec.~\ref{sec:c-inc} below.

Parenthetically, we note that -- at the harmonic level -- one recovers the spin-wave-type
approximation of Kotov {\it et al.}\cite{kotov98} (see Sec.~\ref{sec:bondop2}) if,
instead of calculating $\mu$ and $N_0$ self-consistently, one sets $\mu=-3J'/4$ and $N_0
= 1$.

\section{Triplon interactions}
\label{sec:interaction}

Both the cubic and quartic interaction terms, $\mathcal{H}_3$ \eqref{hcubic2} and
$\mathcal{H}_4$ \eqref{hquartic2}, renormalize the triplon energy and, consequently,
shift the phase boundary of the paramagnetic phase. Moreover, the
cubic term also induces two-particle decay of triplons as discussed in
Refs.~\onlinecite{kole06,zhito06}.

In order to include triplon-triplon interactions into our description,
we treat the effect of $\mathcal{H}_4$ at the mean-field level
(Sec.~\ref{sec:quartic}) and the one of
$\mathcal{H}_3$ perturbatively (Sec.~\ref{sec:cubic}).

\begin{figure}[t]
\centerline{\includegraphics[width=6.4cm]{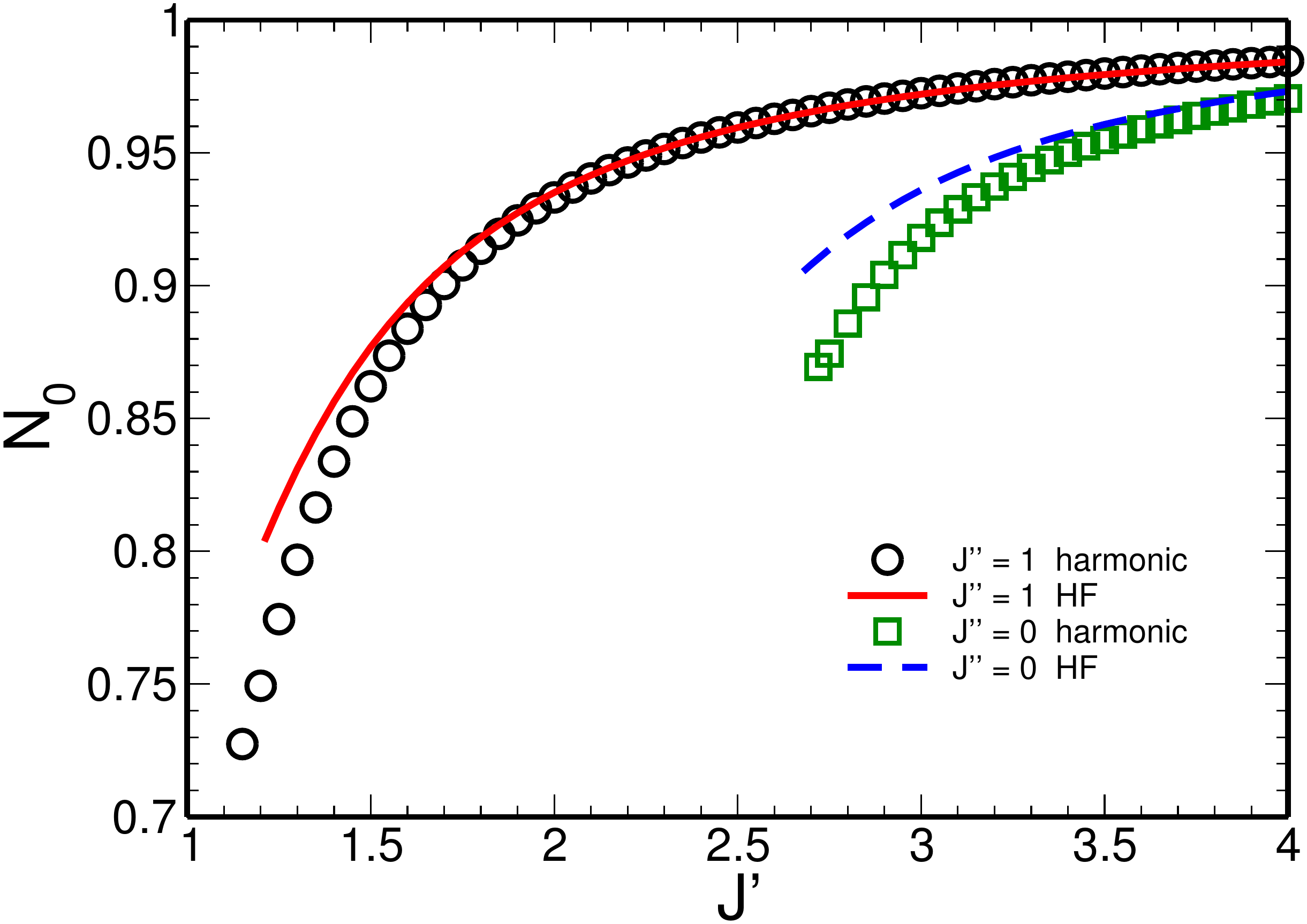}}
\caption{(Color online) Parameter $N_0$ as a function of $J'$ for the
  lines $J'' = 0$ and $J'' = 1$ at the harmonic and Hartree-Fock
  levels.}
\label{fig:nzero}
\end{figure}


\subsection{Hartree-Fock approximation}
\label{sec:quartic}

We treat the quartic triplon interaction within the
self-consistent Hartree-Fock (HF) approximation.
It is straightforward to show that Eq.~\eqref{hquartic2} assumes the
form (for details see Appendix A of Ref.~\onlinecite{cherny09})
\begin{equation}
\mathcal{H}^{\rm HF}_4 = E^{\rm HF}_0 + \sum_\bk  \left[ A^{\rm HF}_\bk
     t^\dagger_{\bk\alpha}t_{\bk\alpha}
   + \frac{1}{2}B^{\rm HF}_\bk \left( t^\dagger_{\bk\alpha}t^\dagger_{\bk\alpha}
                  + {\rm H.c.}\right) \right].
\label{hquartic-hf}
\end{equation}
Here
\begin{eqnarray}
 E^{\rm HF}_0    &=& \frac{3}{N'}\sum_{\bk\,\bq}\gamma_\bk\left(
                   \bar{v}^2_{\bk+\bq}\bar{v}^2_\bq
     - \bar{v}_{\bk+\bq}\bar{u}_{\bk+\bq}\bar{v}_\bq\bar{u}_\bq
  \right),
\nonumber \\
&& \nonumber \\
 A^{\rm HF}_\bk  &=&  -\frac{2}{N'}\sum_\bq\, \gamma_{\bq - \bk}\bar{v}^2_\bq\;
              =   \sum_{n=1}^3 A^{\rm HF}_n\cos(\bk\cdot\taub_n),
\nonumber \\
 &&
\label{hf-coef}\\
 B^{\rm HF}_\bk  &=&  -\frac{2}{N'}\sum_\bq\, \gamma_{\bq - \bk}\bar{u}_\bq\bar{v}_\bq
              =   \sum_{n=1}^3 B^{\rm HF}_n\cos(\bk\cdot\taub_n),
\nonumber
\end{eqnarray}
with $\gamma_\bk$ being the bare quartic vertex \eqref{gamma},
$\bar{u}_\bq$ and  $\bar{v}_\bq$ the Bogoliubov coefficients (see
definition below),
$\taub_n$ the nearest-neighbor vectors \eqref{tauvectors},
and the coefficients $A^{\rm HF}_n$ read
\begin{eqnarray}
  A^{\rm HF}_1 &=& (J'' + 2)\frac{1}{N'}\sum_\bk\cos(k_x)\bar{u}^2_\bk,
\nonumber \\
  A^{\rm HF}_2 &=& \frac{1}{N'}\sum_\bk\cos(\sqrt{3}k_y)\bar{u}^2_\bk,
 \label{hf-coef2} \\
  A^{\rm HF}_3 &=& \frac{1}{N'}\sum_\bk\cos(k_x + \sqrt{3}k_y)\bar{u}^2_\bk.
\nonumber
\end{eqnarray}
Similar expressions hold for $B^{\rm HF}_n$ but with $\bar{u}^2_\bk \rightarrow
\bar{u}_\bk\bar{v}_\bk$.

The final Hamiltonian is now given by $\mathcal{H} = E_0 +
E^{\rm HF}_0 + \mathcal{H}_2 + \mathcal{H}^{\rm HF}_4$.
It is quadratic in triplet operators and can also be
diagonalized by the Bogoliubov transformation \eqref{bogo-transf}, namely
\begin{equation}
\mathcal{H} = \bar{E}^{\rm HF}_0
             + \sum_{\bk\,\alpha} \bar{\omega}_\bk
                 b^\dagger_{\bk\alpha}b_{\bk\alpha}.
\label{diag-ham2}
\end{equation}
In the above expression, the renormalized triplon energy
$\bar{\omega}_\bk$ is equal to Eq.~\eqref{omega-harmonic} but
now $A_\bk \rightarrow \bar{A}_\bk = A_\bk + A^{\rm HF}_\bk$ and
$B_\bk \rightarrow \bar{B}_\bk = B_\bk + B^{\rm HF}_\bk$.
The ground state energy $\bar{E}^{\rm HF}_0$ is similar to
Eq.~\eqref{egs-harmonic} apart from the replacements $\omega_\bk
\rightarrow \bar{\omega}_\bk$ and $A_\bk \rightarrow \bar{A}_\bk$
and the inclusion of $E^{\rm HF}_0$.
Finally, the Bogoliubov coefficients and the analog of the
self-consistent equations \eqref{self-mu} and \eqref{self-nzero} now
read
\begin{eqnarray}
 \bar{u}^2_\bk , \bar{v}^2_\bk &=& \pm 1/2 + \bar{A}_\bk/2\bar{\omega}_\bk,
 \;\;\;\;\;
 \bar{u}_\bk\bar{v}_\bk = \bar{B}_\bk/2\bar{\omega}_\bk,
\label{bogo-coef-hf} \\
&& \nonumber \\
   \mu &=& -\frac{3J'}{4} + \frac{3}{N_0}\sum_\bk B_\bk \bar{v}_\bk
           \left(\bar{v}_\bk - \bar{u}_\bk \right),
\label{self-mu-hf} \\
&& \nonumber \\
   N_0 &=&  1 - \frac{3}{N'}\sum_\bk \bar{v}^2_\bk.
\label{self-nzero-hf}
\end{eqnarray}
Note that $B_\bk$ (not $\bar{B}_\bk$) enters the momentum summation
in Eq.~\eqref{self-mu-hf} (for details, see Appendix \ref{ap:details2}).
In addition to $\mu$ and $N_0$, now $A^{\rm HF}_{1,2,3}$ and $B^{\rm HF}_{1,2,3}$ are
also self-consistently calculated. The set of equations
\eqref{hquartic-hf}--\eqref{self-nzero-hf} constitutes HF approximation.

The resulting triplon dispersion is included in Fig.~\ref{fig:disp-harmonic}, which shows
that corrections arising from $\mathcal{H}^{\rm HF}_4$ are small, except for a slight
upward renormalization of the gap. For a fixed $J'$, the minimum wavevector $\bQ_0$
displays the same qualitative behavior in terms of $J''$ as found in the harmonic
approximation, see Fig.~\ref{fig:min01}. In the incommensurate regime, $\bQ_0$
numerically deviates from the harmonic result for small $J'$ (but recovers the harmonic
results in the large-$J'$ limit), see Fig.~\ref{fig:min}. Moreover, the line marking the
CIT is slightly shifted as well, see Fig.~\ref{fig:min01}.

Fig.~\ref{fig:nzero} shows the evolution of $N_0$ as a function of $J'$ for fixed
$J''$, again comparing the HF results to those from the harmonic
approximation. The deviations are small in general and vanish in the strong-dimer limit
$J'\to\infty$. Inside the disordered phase, we observe $N_0 > 0.7$ which implies that
$v_\bk$ and $\bar{v}_\bk \lesssim 0.1$, see Eqs.~\eqref{self-nzero} and
\eqref{self-nzero-hf}. The fact that $\bar{v}_\bk$ is a small quantity
is used in order to derive the self-consistent equations
\eqref{self-mu-hf} and \eqref{self-nzero-hf}, see Appendix
\ref{ap:details2} for details.

We note that we experienced more difficulties in obtaining self-consistent
solutions near the phase boundary in the HF approximation as compared to the
harmonic one, but these problems are cured upon including the cubic term, see
Sec.~\ref{sec:cubic} below.

\begin{figure}[t]
\centerline{\includegraphics[width=6.4cm]{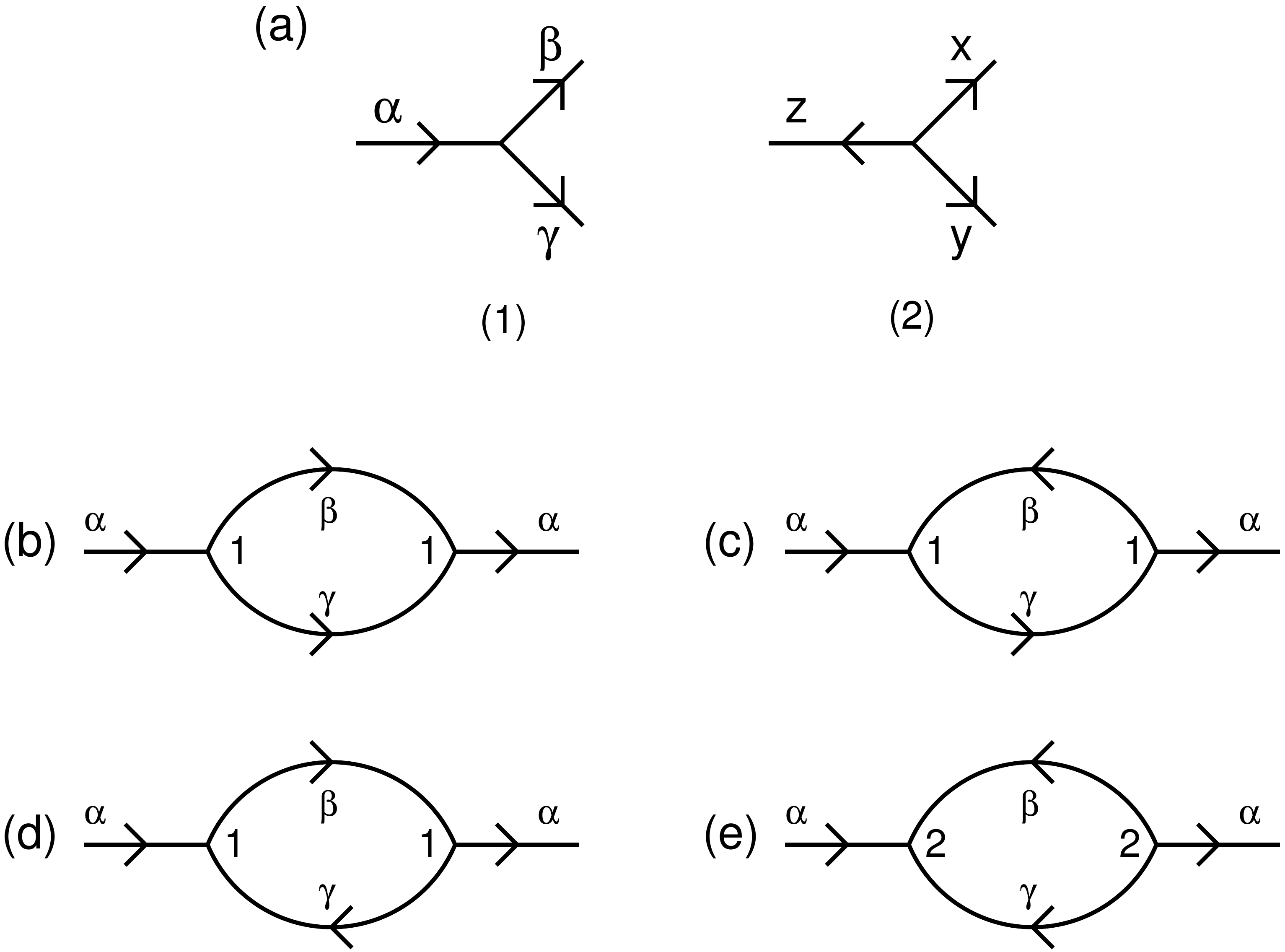}}
\caption{(a) Renormalized cubic vertices $\Gamma_1(\bk,\bp)$, (1), and
         $\Gamma_2(\bk,\bp)$, (2). Here $(\alpha,\beta,\gamma)$ = $(x,y,z)$,
         $(z,x,y)$, and $(y,z,x)$.
         (b)--(e) Lowest-order diagrams resulting
         from the combination of vertices (1) and (2) which contribute to
         the normal self-energy $\Sigma_3(\bk,\omega)$.}
\label{fig:cubic-diag}
\end{figure}


\subsection{Hartree-Fock-cubic approximation}
\label{sec:cubic}

We implement a Hartree-Fock-cubic (HF-cubic) approximation by perturbatively adding the
effect of $\mathcal{H}_3$ to the mean-field Hamiltonian $\mathcal{H} = E_0 + E^{\rm HF}_0 +
\mathcal{H}_2 + \mathcal{H}^{\rm HF}_4$. This scheme, which considerably simplifies the
treatment of the cubic triplet interaction, is similar in spirit to the one adopted in
Refs.~\onlinecite{kotov98,kotov99b}, where an interacting Hamiltonian for triplets without
the cubic term is treated self-consistently first, and then the lowest-order corrections
due to the cubic term are added.

\begin{figure*}[t]
\centerline{\includegraphics[height=6.1cm]{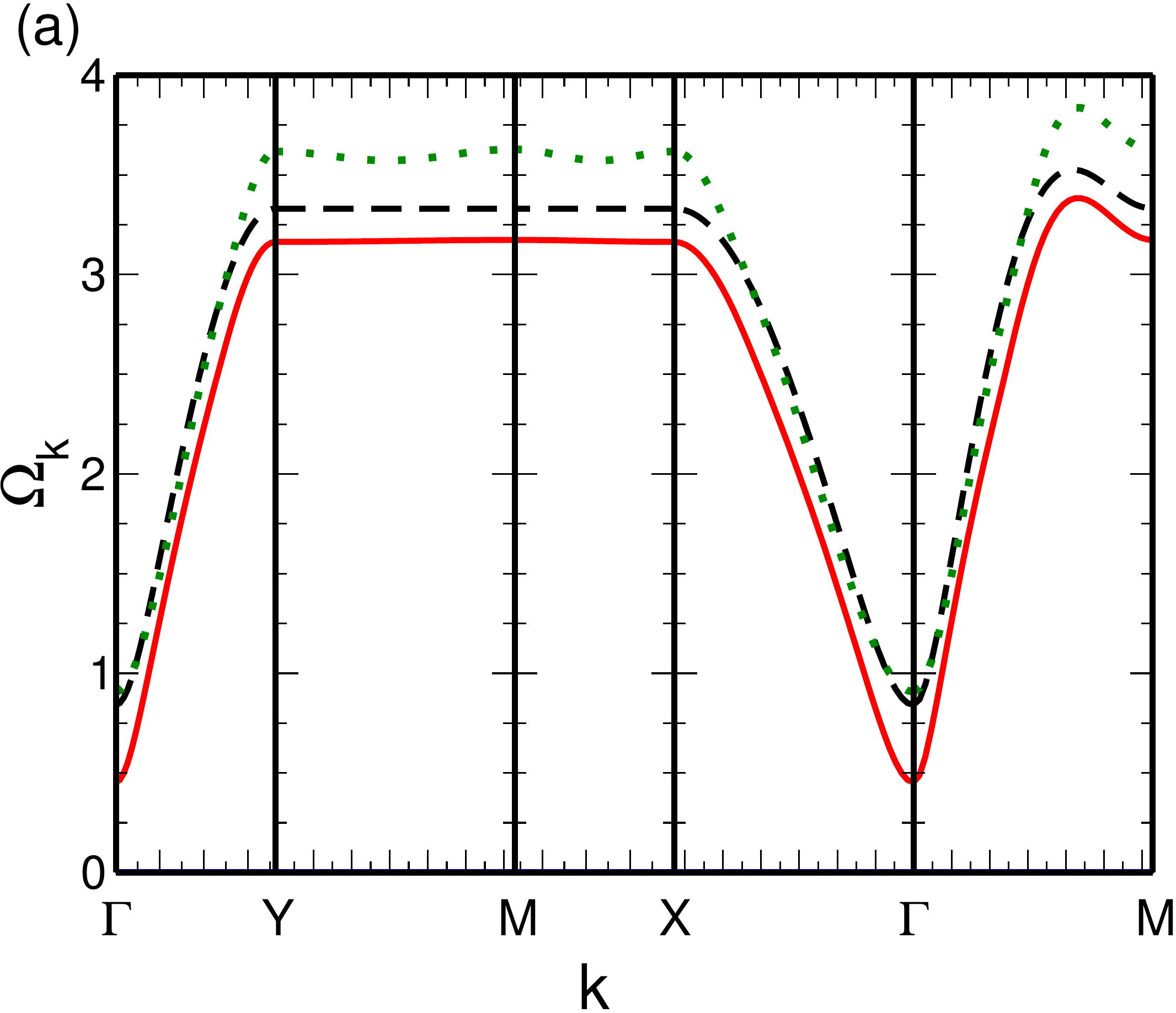}
 \hskip0.5cm
 \includegraphics[height=6.1cm]{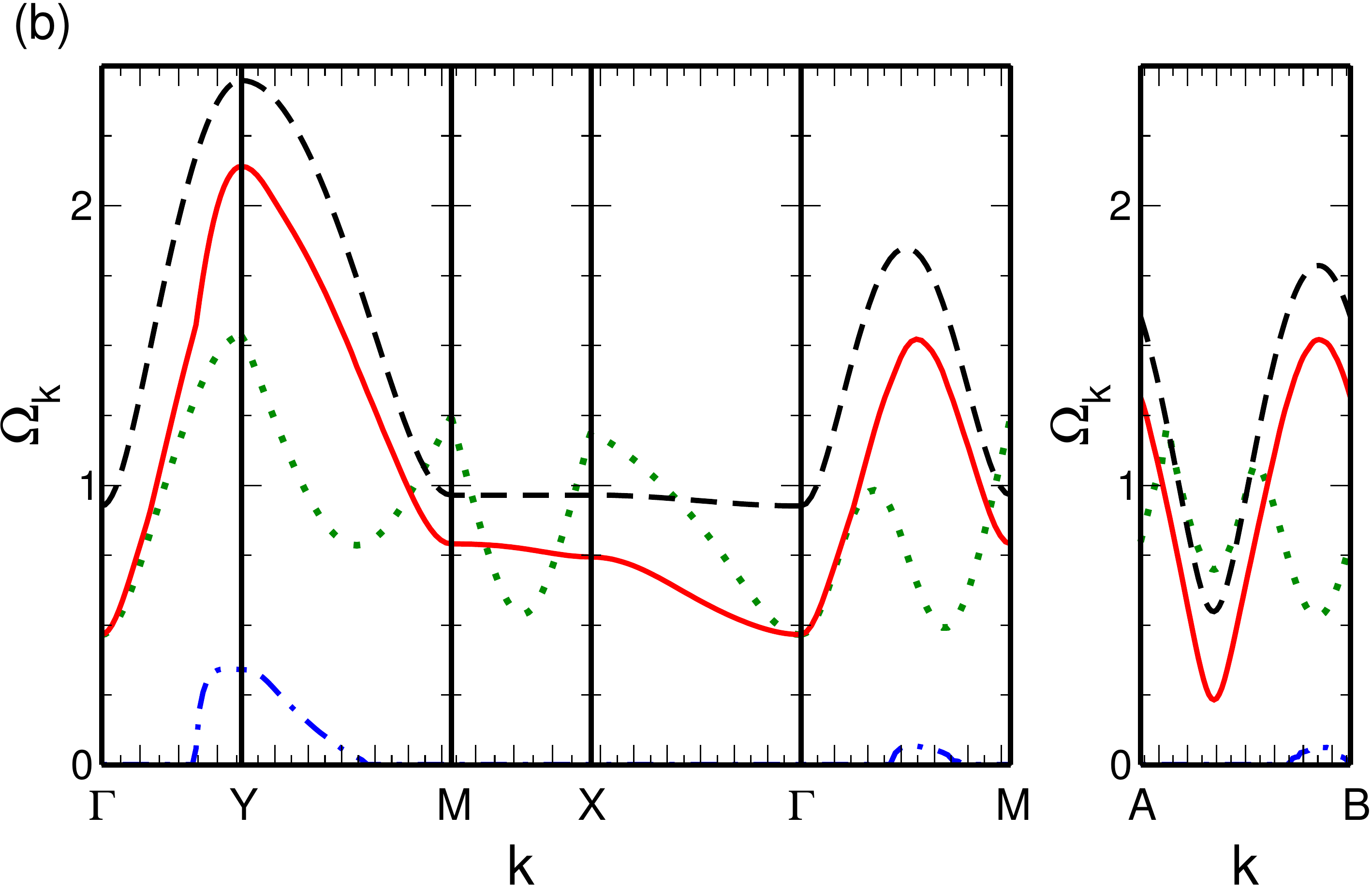}}
\caption{(Color online) Triplon dispersion relation along some
         particular lines of the dimerized Brillouin zone
         for (a) $J' = 3$ and $J'' = 0$ and (b) $J' = 1.5$ and
         $J'' = 1$ at the HF (dashed black line)
         and HF-cubic (solid red line) approximations. The triplon decay rate
         $\tilde{\Gamma}_\bk$ (dot-dashed blue line) and the bottom of the
         two-particle continuum (dotted green line) both at the HF-cubic
         approximation are also shown. The A--B line includes the
         triplon minimum energy of the corresponding approximation.
}
\label{fig:disp}
\end{figure*}

Using the Bogoliubov transformation \eqref{bogo-transf} with the
renormalized coefficients $\bar{u}_\bk$ and $\bar{v}_\bk$
\eqref{bogo-coef-hf}, one can show that $\mathcal{H}_3$
\eqref{hcubic2} in terms of the $b^\dagger_{\bk\alpha}$ and
$b_{\bk\alpha}$ operators reads
\begin{eqnarray}
\mathcal{H}_3 &=& \frac{1}{2\sqrt{N'}}\sum_{\bk,\bp}
                   \sideset{}{'}\sum_{\alpha,\beta,\gamma}
                   \Gamma_1(\bk,\bp)
               (b^\dagger_{\bk-\bp \alpha}b^\dagger_{\bp \beta}b_{\bk \gamma}
                       + {\rm H.c.})
\nonumber \\
             &+& \frac{1}{2\sqrt{N'}}\sum_{\bk,\bp}
                 \Gamma_2(\bk,\bp)
                   (b^\dagger_{\bk-\bp x}b^\dagger_{\bp y}b^\dagger_{-\bk z}
                 + {\rm H.c.}).
\end{eqnarray}
Here the sum over $\alpha,\beta,\gamma$ has only three components:
$(\alpha,\beta,\gamma) = (x,y,z)$, $(z,x,y)$, and $(y,z,x)$. The
renormalized vertex $\Gamma_1(\bk,\bp)$ is given by
\begin{eqnarray}
\Gamma_1(\bk,\bp) &=&
   \left(\xi_{\bk-\bp} - \xi_\bp\right)\left(\bar{u}_{\bk-\bp}\bar{u}_\bp \bar{u}_\bk
                      +  \bar{v}_{\bk-\bp}\bar{v}_\bp \bar{v}_\bk \right)
\nonumber \\
        &+& \left(\xi_\bk + \xi_\bp\right)\left(\bar{v}_{\bk-\bp}\bar{u}_\bp \bar{v}_\bk
                      +  \bar{u}_{\bk-\bp}\bar{v}_\bp \bar{u}_\bk \right)
\nonumber \\
        &-&  \left(\xi_{\bk-\bp} + \xi_\bk\right)\left(\bar{v}_{\bk-\bp}\bar{u}_\bp \bar{u}_\bk
                      +  \bar{u}_{\bk-\bp}\bar{v}_\bp \bar{v}_\bk
                      \right),\;\;\;\;
\label{gammas}
\end{eqnarray}
with $\xi_\bp$ being the bare cubic vertex \eqref{xi}, and the vertex
$\Gamma_2(\bk,\bp) = - \Gamma_1(\bk,\bp)$ in addition to the
replacements $\bar{u}_\bk \leftrightarrow \bar{v}_\bk$. The vertices
$\Gamma_1(\bk,\bp)$ and $\Gamma_2(\bk,\bp)$ are illustrated in
Fig.~\ref{fig:cubic-diag}a.

The lowest-order diagrams which contribute to the (normal) self-energy
$\Sigma_3(\bk,\omega)$ are shown in Figs.~\ref{fig:cubic-diag}(b)--(e). In each diagram, the
solid line corresponds to the bare $b$ triplon propagator (we now omit the index $\alpha$
since the three triplon branches are degenerate)
\begin{equation}
 G^{-1}_0(\bk,\omega) = \omega - \bar{\omega}_\bk + i\delta.
\label{bare-propag}
\end{equation}
Note here that no anomalous bare $b$ propagators exist; in principle, those are
generated in perturbation theory, but will be neglected in the following.\cite{comment01}
Using standard diagrammatic techniques for bosons at zero temperature,
one shows that only the diagrams (b) and (e) are finite, and therefore
$\Sigma_3(\bk,\omega) = \Sigma^{(b)}_3(\bk,\omega) +
\Sigma^{(e)}_3(\bk,\omega)$ with
\begin{eqnarray}
   \Sigma^{(b)}_3(\bk,\omega) &=& \frac{1}{4N'}\sum_\bq
      \frac{\Gamma^2_1(\bk,\bq)}{\omega - \bar{\omega}_\bq
                       - \bar{\omega}_{\bk-\bq} + i\delta},
\nonumber \\
    \Sigma^{(e)}_3(\bk,\omega) &=& -\frac{1}{4N'}\sum_\bq
       \frac{\Gamma^2_2(\bk,\bq)}{\omega + \bar{\omega}_\bq
                       + \bar{\omega}_{\bk-\bq} - i\delta}.
\label{selfenergy-cubic2}
\end{eqnarray}

The renormalized triplon energy $\Omega_\bk$ is given by the poles of
the full Green's function $G(\bk,\omega)$, i.e,
\begin{equation}
 G^{-1}(\bk,\omega) = \omega - \bar{\omega}_\bk - \Sigma_3(\bk,\omega) = 0.
\label{poles}
\end{equation}
We solve Eq.~\eqref{poles} within the so-called off-shell
approximation, which consists in evaluating the self-energy at $\omega = \Omega_\bk -
i\tilde{\Gamma}_\bk$, i.e.
\begin{equation}
   \Omega_\bk - i\tilde{\Gamma}_\bk  - \bar{\omega}_\bk
              - \Sigma_3(\bk,\Omega_\bk - i\tilde{\Gamma}_\bk) = 0.
\label{poles2}
\end{equation}
[Recall that in the more common on-shell approximation, the
self-energy is evaluated at the bare single-particle energy, i.e.,
$
   \Omega_\bk - i\tilde{\Gamma}_\bk  - \bar{\omega}_\bk
              - \Sigma_3(\bk,\bar{\omega}_\bk) = 0.
$]
Moreover, instead of using Eq.~\eqref{poles2}, causality requires to
consider
\begin{equation}
   \Omega_\bk - i\tilde{\Gamma}_\bk  - \bar{\omega}_\bk
              - \Sigma_3(\bk,\Omega_\bk + i\tilde{\Gamma}_\bk) = 0.
\label{poles3}
\end{equation}

The procedure outlined above follows Ref.~\onlinecite{cherny09} where spin-wave
excitations of the isotropic triangular-lattice AF Heisenberg model
($J' = J'' = 1$) were calculated. It is shown by Chernyshev {\it et
  al.}\cite{cherny09} that the on-shell approximation leads
to discontinuities in the spin-wave spectrum and concomitant
logarithmic singularities in the decay rate $\tilde{\Gamma}_\bk$, and
that the off-shell approximation regularizes such singularities.
Finally, the replacement $\Omega_\bk - i\tilde{\Gamma}_\bk \rightarrow
\Omega_\bk + i\tilde{\Gamma}_\bk$ in the argument of the self-energy
guarantees that the quasiparticle pole is in the correct half of the complex plane
(we refer the reader to the Appendix D, Ref.~\onlinecite{cherny09},
for details).

In Fig.~\ref{fig:disp}, we compare the renormalized triplon energy
$\Omega_\bk$ obtained in the HF--cubic approximation to that from
HF. Clearly, the inclusion of the cubic term leads to sizeable changes
of the dispersion, which are particularly pronounced in the
incommensurate regime [see Fig.~\ref{fig:disp}~b for $J' = 1.5$ and
$J'' = 1$]. For instance, the dispersion along ${\bf\Gamma}-{\bf M}$ is
significantly enhanced by the cubic term.
Most notably, the triplon gap $\Delta$ is renormalized {\em downwards},
see Fig.~\ref{fig:gap-egs} below. In other words, cubic interactions tend to
destabilize the paramagnetic phase, whereas (repulsive) quartic
interactions have the opposite effect.

The behavior of the minimum wavevector $\bQ_0$ as a function of $J''$
is again qualitatively similar to the harmonic one,
Fig.~\ref{fig:min01}. Interestingly, the shift of the CIT line due to the cubic
interaction is opposite (and larger) to that induced by the quartic interaction, such
that the CIT line is now located at some $J''_{\rm CIT} > 0.75$ which moreover depends
non-monotonically on $J'$, Fig.~\ref{fig:phase-diag}.

Fig.~\ref{fig:disp} also displays the decay rate of triplons caused by
two-particle decay via $\mathcal{H}_3$ --- this is non-zero whenever
the single-particle branch is above the two-particle threshold.

We conclude that the cubic triplon term is of vital importance for a
quantitative description of the triplon dynamics in the frustrated
coupled-dimer model \eqref{ham}, especially in the incommensurate
regime. This is in contrast to, e.g., the unfrustrated
asymmetric bilayer model studied in Ref.~\onlinecite{kotov98} where
the cubic term leads to minor corrections only. A further discussion
of physical implications of our results is given below.


\section{Discussion}
\label{sec:discussion}

\subsection{Phase boundary}

\begin{figure*}[!t]
\centerline{
  \includegraphics[height=4.5cm]{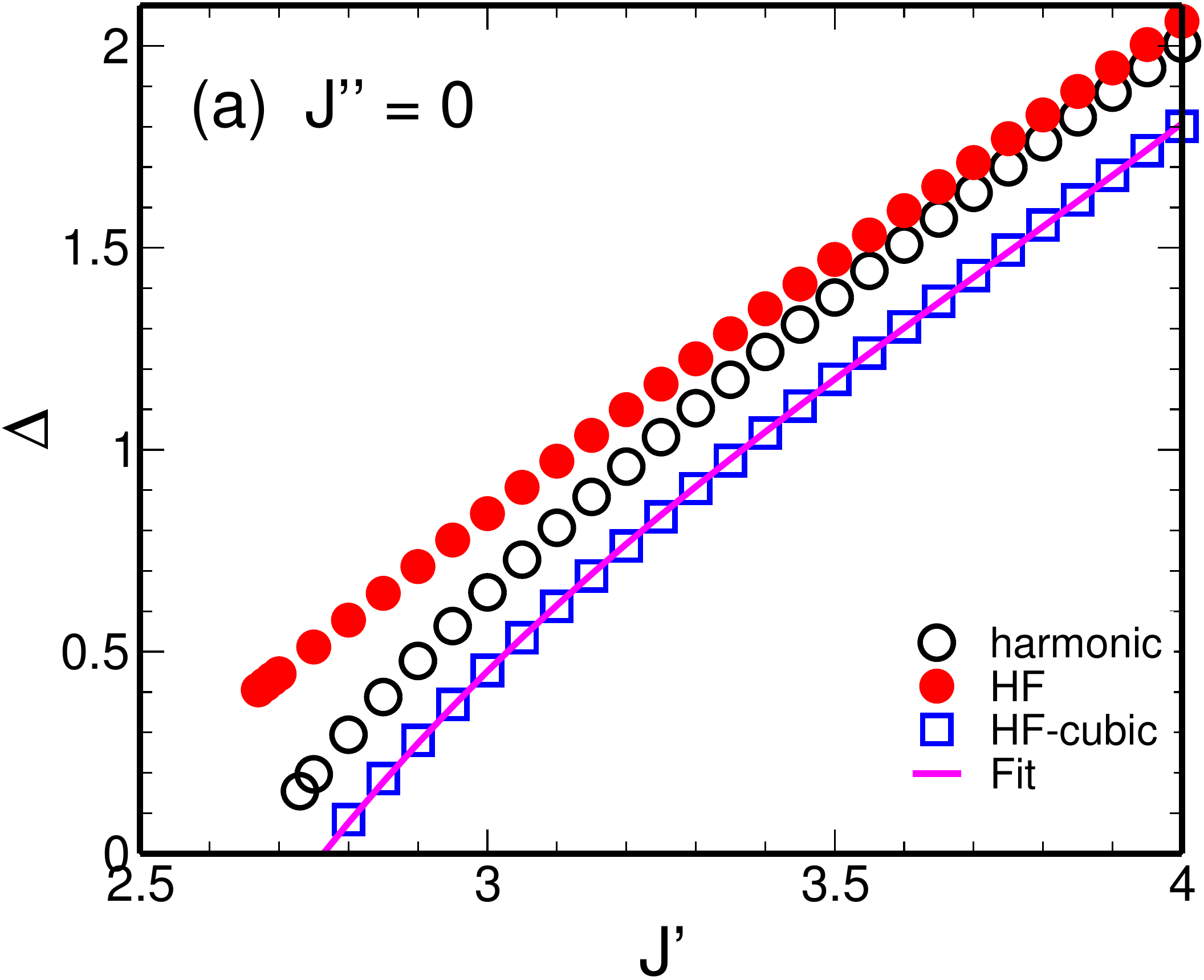}
  \hskip0.15cm
  \includegraphics[height=4.5cm]{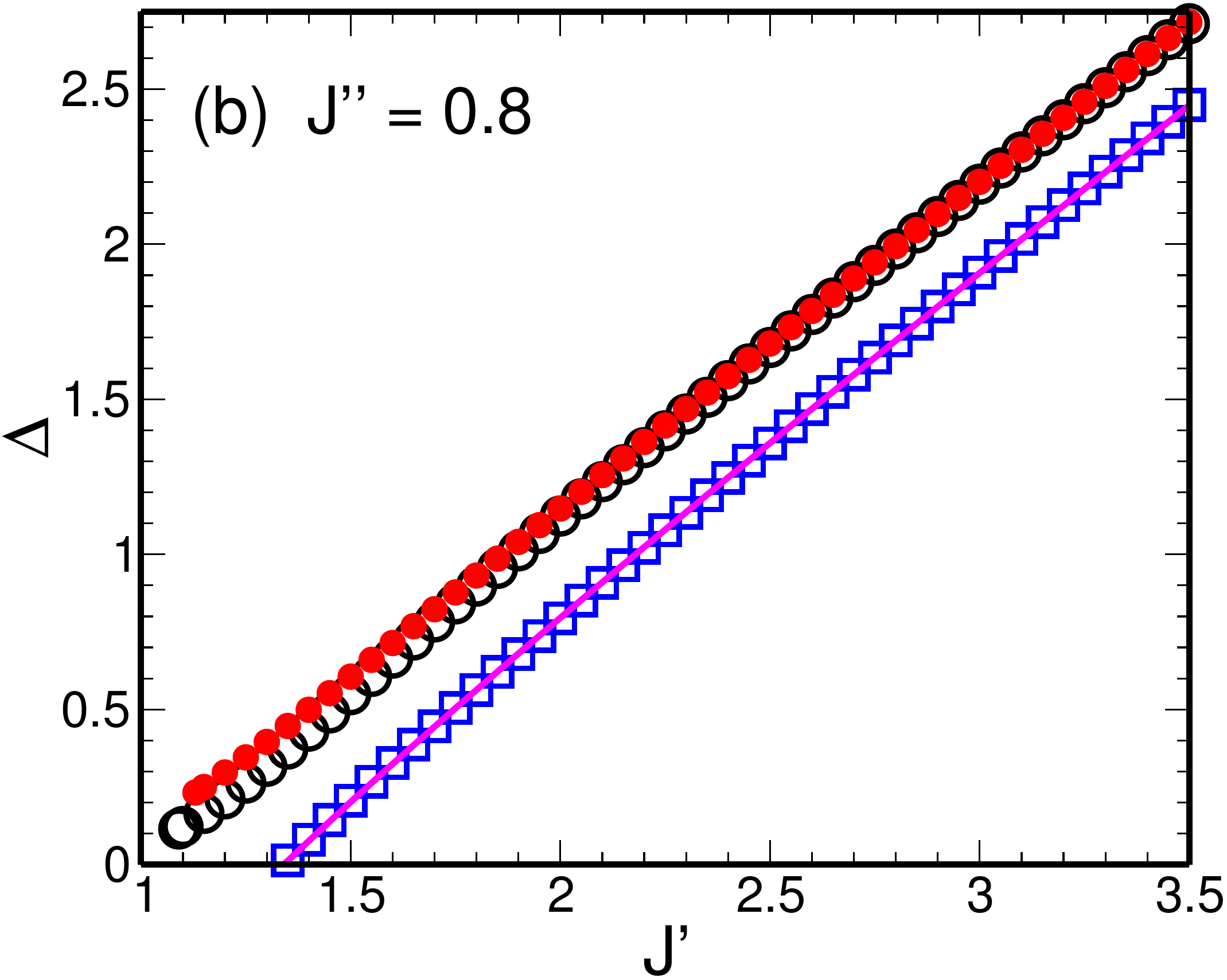}
  \hskip0.15cm
  \includegraphics[height=4.5cm]{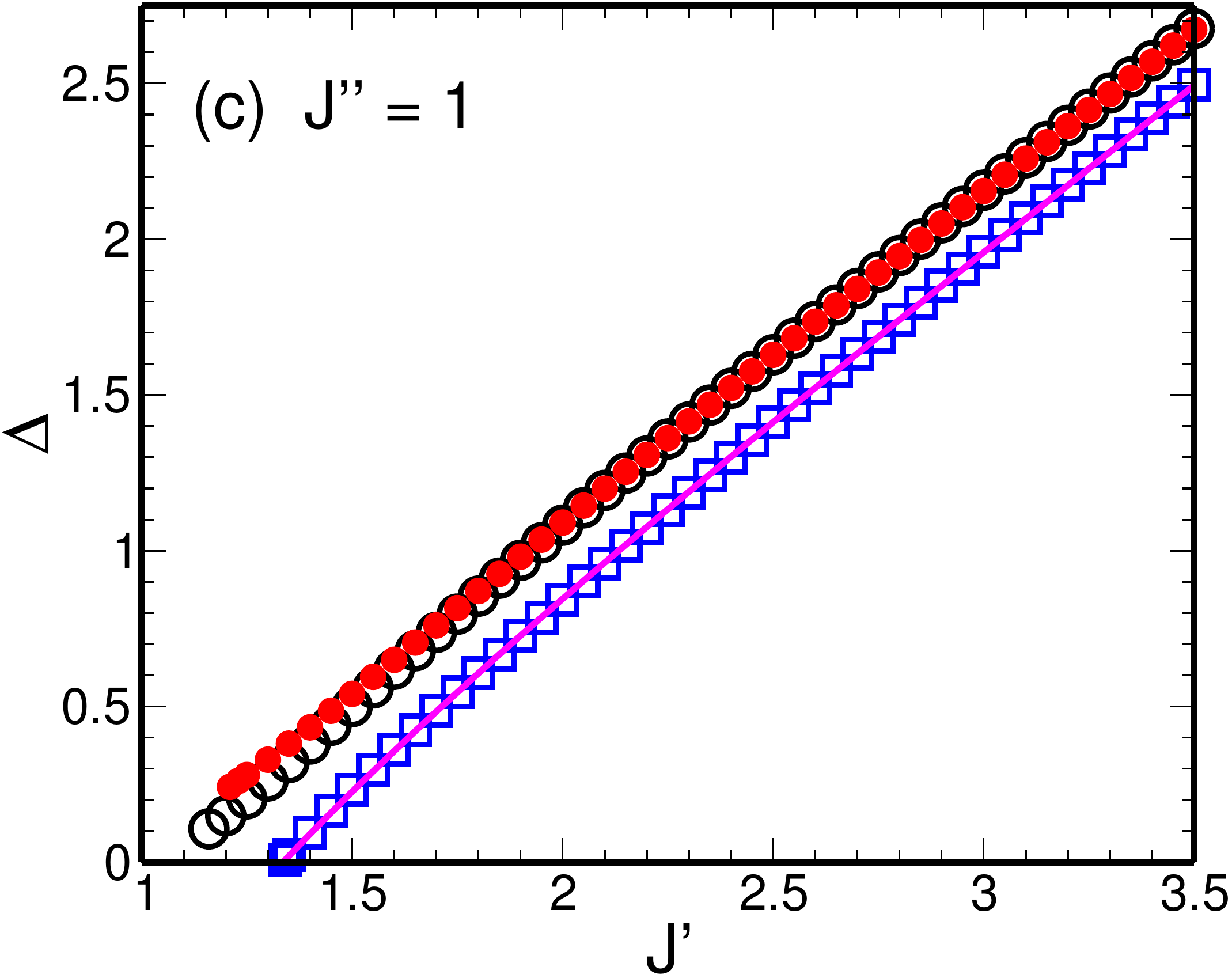}
}
\caption{(Color online) Triplon gap $\Delta$ as a function of $J'$ for
         (a) $J'' = 0$, (b) $J'' = 0.8$, and (c) $J'' = 1$ at
         the harmonic ($\Circle$), HF (${\color{red}\CIRCLE}$),
         and HF-cubic (${\color{blue}\square}$)
         approximations. The solid line is a fit to the data using
         Eq.~\eqref{fit}.
}
\label{fig:gap-egs}
\end{figure*}

For the different levels of approximation, the evolution of the energy gap $\Delta$ with
the dimerization strength $J'$ is shown in Fig.~\ref{fig:gap-egs} for
$J''=0$, $0.8$, and $1$.
As expected, $\Delta$ increases with $J'$; the vanishing of the gap defines a critical
value $J'_c$ where the singlet phase becomes unstable towards magnetic order. Assuming a
continuous QPT, we fit the data to
\begin{equation}
\Delta = a_0 + a_1J' + a_2\frac{1}{J'} + a_3(J')^2,
\label{fit}
\end{equation}
and use the condition $\Delta = 0$ to estimate the critical coupling $J'_c$. (Note that
the present approximations cannot be expected to yield non-trivial critical exponents.
Omitting the $1/J'$ fitting term leads to only minor changes of $J'_c$. )

The resulting phase boundary for $0<J''<1.5$ is shown in Fig.~\ref{fig:phase-diag}, with
the HF-cubic result (${\color{red}\square}$) being our best approximation. The phase
diagram in Fig.~\ref{fig:phase-diag} displays four distinct regions: At large $J'$ we
have the gapped dimer phase, with spin correlations peaked at $\bQ_0 = (0,0)$ for $0 \le
J'' \le J''_{\rm CIT}$ and incommensurate $\bQ_0$ for $J''>J''_{\rm CIT}$. At the
harmonic level $J''_{\rm CIT}$ equals $0.75$, but acquires a $J'$
dependence from anharmonic terms, see Figs.~\ref{fig:min01} and \ref{fig:min}.
The closing of the triplon gap at $J'_c$ for a given $J''$ leads to magnetic long-range
order at wavevector $\bQ_0$ (provided that no other phase intervenes). One concludes that
the system displays collinear N\'eel order for $J < J''_{\rm CIT}$ and non-collinear
spiral order for $J''> J''_{\rm CIT}$. As indicated in
Fig.~\ref{fig:schem}, an additional paramagnetic phase
might be realized near the crossing of the CIT and the order-disorder transition
lines,\cite{series-exp} with a more detailed discussion given in Sec.~\ref{sec:c-inc}.

Let us quantitatively discuss the evolution for two values of $J''$.
(i) $J'' = 0$: Here we find a critical coupling $J'_c = 2.65$ ($2.76$) at the
harmonic (HF-cubic) level, which is quite close to the one determined via QMC
simulations for the (topologically equivalent) staggered dimerized AF
Heisenberg model on a square lattice, ${J'}_c^{\rm QMC} = 2.5196$.\cite{wenzel08}
For $J' < J'_c$ the system indeed develops collinear
N\'eel order with $\bQ_0=(0,0)$.\cite{wenzel08,stag-dimer}
(ii) $J'' = 1$, where we find $J'_c = 1.09$ ($1.34$) at the harmonic
(HF-cubic) level. Again, this is a very
reasonable result since LRO is certainly present at the isotropic point ($J' = J'' = 1$).
One difference to case (i) is that the ordering wavevector varies with $J'$ inside the
ordered phase (see appendix~\ref{ap:class} for the corresponding classical-limit results);
therefore, $\bQ_0$ at the phase boundary is distinct from the Goldstone wavevectors ${\bf
K},{\bf K'}$ of the $120^\circ$ structure, see also Fig.~\ref{fig:model}c.

\subsection{Commensurate--incommensurate transition}
\label{sec:c-inc}

As shown above, the minimum wavevector $\bQ_0$ of the triplon
dispersion in the gapped paramagnetic phase of the Heisenberg model
\eqref{ham} is locked to $(0,0)$ for small $J''$, while it moves to
incommensurate values for larger $J''$ (Figs.~\ref{fig:min01} and \ref{fig:min}),
with the boundary being located near $J''_{\rm CIT} \lesssim 0.75$. This
commensurate-incommensurate transition (CIT), driven by increasing
magnetic frustration, has various consequences.

\begin{figure}[t]
\vskip-0.85cm
\centerline{\includegraphics[width=8.1cm]{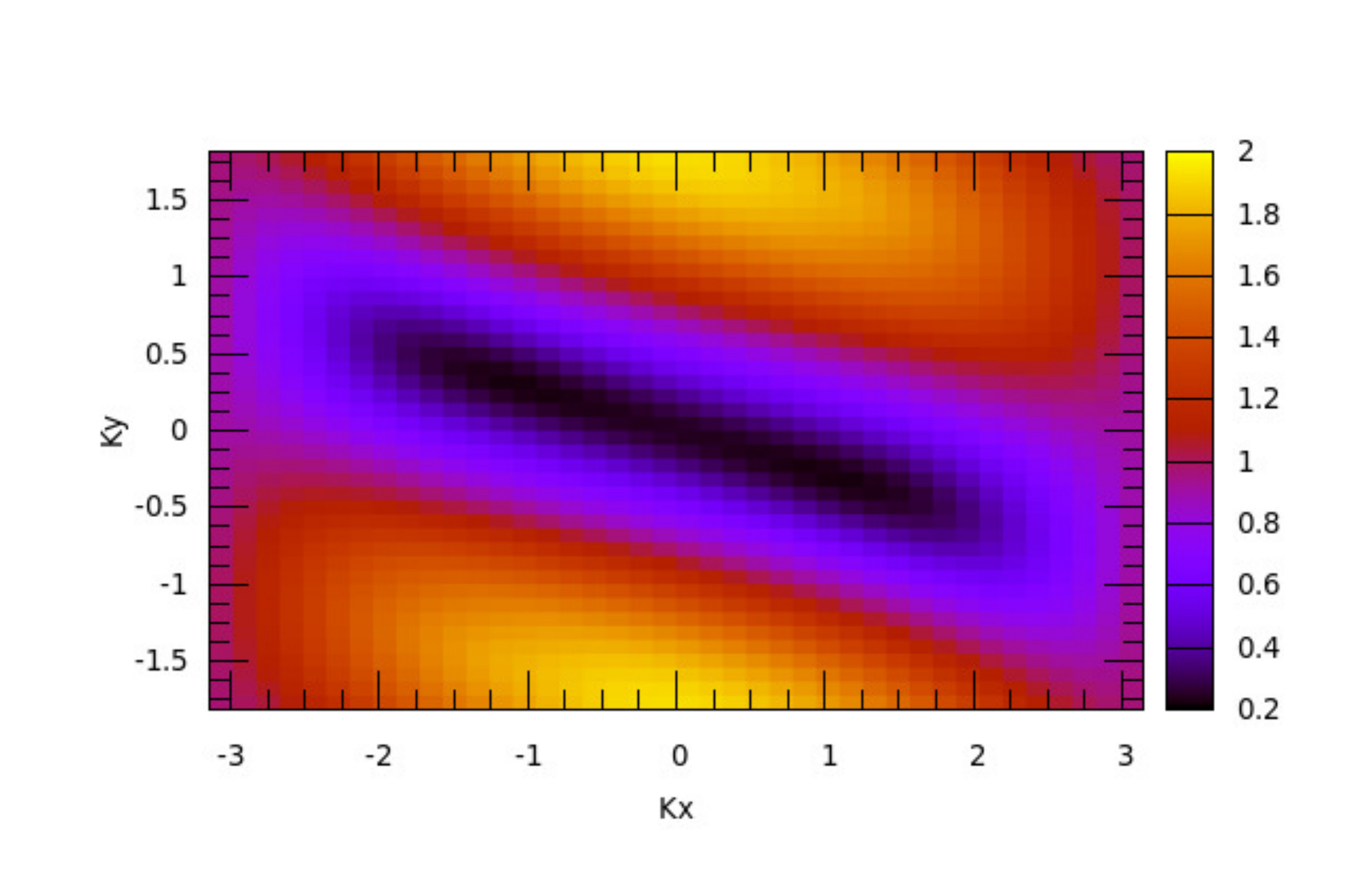}}
\vskip-0.1cm
\caption{(Color online) Contour plot of the triplon energy
  $\Omega_\bk$ (HF-cubic), for $J' = 1.5$
  and $J'' = 0.8$, i.e., near the CIT.}
\label{fig:disp-c-ic}
\end{figure}

Right at the CIT, the quadratic piece of the triplon dispersion vanishes in one of the
two space directions (independent of the level of approximation), i.e., we find for the
triplon propagator
\begin{equation}
 G^{-1}(\bk,\omega) = -\omega^2 + \Delta^2 + c_1\tilde{k}_1^2
                            + d_1 \tilde{k}_2^4
 \label{glif}
\end{equation}
in a small-momentum expansion around the ${\bf \Gamma}$ point, where
$\tilde{k}_{1,2}$  are the two components of $\bk$
perpendicular and parallel to the $\bQ_0$ which is taken beyond the CIT.
Moreover, in the incommensurate regime near the CIT the dispersion
along the $\tilde{k}_2$ direction (connecting ${\bf\Gamma}$ and $\bQ_0$) is
anomalously flat. This is illustrated in Fig.~\ref{fig:disp-c-ic},
which shows a contour plot of the triplon
dispersion $\Omega_\bk$ (HF-cubic approximation) for
$J' = 1.5$ and $J'' = 0.8$.
Qualitatively, the soft dispersion near the CIT will lead to an
effective dimensional reduction. When the triplon gap closes,
$\Delta=0$, Eq.~\eqref{glif} defines a quantum
Lifshitz point, see Sec.~\ref{sec:qcrit} below.

Of course, the physics underlying the CIT is also observed inside the {\em ordered}
phase. This has been studied in particular in the non-dimerized case, $J'=1$.
Linear spin-wave theory\cite{trumper99,merino99}
yields a QPT from collinear N\'eel to spiral order at
$J''_c = 0.5$, where the ordering wavevector changes from a
commensurate to an incommensurate value.
(The analysis of the classical ground state for the general case
$J'\neq 1$ is given in Appendix \ref{ap:class}.)
The magnetization curve has a minimum at the transition, where the
spin-wave velocity vanishes along one particular direction in $k$
space. Coupled-cluster calculations\cite{ccm} find a scenario similar
to spin-wave theory, but here $J''_c \approx 0.8$. Finally,
series expansion results\cite{series-exp} not only indicate a CIT at
$J'' \approx 0.75$, but they also provide evidence for a disordered
phase in the vicinity of the CIT,
i.e, for $0.68 < J'' < 0.91$.


\subsection{Two-particle decay}

The cubic triplon term $\mathcal{H}_3$ is generically present in the model \eqref{ham}
and thus enables two-particle decay of triplons.\cite{kole06,zhito06} Such decay occurs
if the one-triplon dispersion moves inside the two-particle continuum.

As can be seen from Fig.~\ref{fig:disp}, two-particle decay does not happen in the
commensurate regime of small $J''$, where one-particle spectrum is always below the
two-particle continuum. Note that an exceptional case is realized right at the QPT to the
magnetically ordered state, where the lower bound of the two-particle continuum coincides
with the single-particle branch near the ordering wavevector $(0,0)$. This non-trivial
coupling has been argued to lead to a novel universality class for the phase
transition,\cite{stag-dimer} see also Sec.~\ref{sec:qcrit} below.

The situation is different in the incommensurate regime, where high-energy triplons can
decay into triplon pairs, see Fig.~\ref{fig:disp} for the calculated decay rate. Note
that such decay never happens near $\bQ_0$, i.e., cubic interactions are not part of the
critical theory in the incommensurate case.

As an aside, we note that decay of high-energy modes also occurs inside the ordered
phase, where spin waves can decay into pairs of spin waves provided that the order is
non-collinear, for details see Ref.~\onlinecite{cherny09}.


\subsection{Quantum criticality}
\label{sec:qcrit}

Let us briefly discuss the quantum phase transitions in the phase
diagram shown in Fig.~\ref{fig:schem}.

The transition from the dimer phase to the collinear LRO state with
$\bQ_0=(0,0)$ would in principle be expected to be in the standard
Heisenberg [or O(3)] universality class in $D=2+1$ dimensions;
however, the cubic triplon term becomes part of the critical theory
and leads to a new universality class (labelled class B in
Ref.~\onlinecite{stag-dimer}), with leading O(3) exponents and
anomalously large corrections to scaling.\cite{wenzel08,stag-dimer}

The transition from the dimer phase to the non-collinear LRO state
does not display this complication and therefore is a standard O($N$)
QPT in $D=2+1$, with $N=4$ (assuming a single transition
to coplanar spiral order).

Finally, we can discuss the multicritical point where the collinear
LRO, non-collinear LRO, and
dimer phases meet. Here, the softness of the triplon dispersion in one
direction, Eq.~\eqref{glif}, implies that the quadratic piece of the
critical field theory takes the form
\begin{equation}
\mathcal{S} = \int d^2 x d\tau\, \phi_a (\delta + \partial_\tau^2 + \partial_x^2 +
\partial_y^4) \phi_a
\label{slif}
\end{equation}
where $\phi_a$ are the components of the magnetic order parameter, and $\delta=0$ defines
the phase transition point. Such $(d,m)$ quantum Lifshitz points (here $d=2$, and $m=1$
refers to the number of ``soft'' directions) have been discussed
before,\cite{qlif1,qlif2} but the relevant case of $d=2$ dimensions with undamped
order-parameter dynamics has not been studied in any detail. A thorough analysis of this
critical theory is therefore deferred to a future publication; here we only make a few
qualitative remarks.
The absence of the quadratic derivative in $y$ direction implies that the effective
dimensionality is reduced compared to a standard $\phi^4$ theory, or in other words, the
upper critical dimension (in the quantum case) is increased from $d_c^+=3$ to
$d_c^+=3+m/2$ (Ref.~\onlinecite{qlif2}).
While Eq.~\eqref{slif}, supplemented by a standard quartic term (note that a cubic
term\cite{stag-dimer} might be important as well), describes a continuous multicritical
point, it is conceivable that this is pre-empted by a fluctuation-induced transition into
a novel phase. We speculate that this could be a non-trivial paramagnetic phase (e.g.
with a symmetry-breaking dimerization), as deduced for $J'=1$ from series-expansion
studies.\cite{series-exp}


\subsection{Application to \pdmit}
\label{sec:exp}

As mentioned in the Introduction, the experimental
findings\cite{dmit01,dmit02,dmit03,dmit_rev} indicate that the organic Mott insulator
\pdmit\ realizes a columnar VBS phase at low temperature and pressure. According to
Ref.~\onlinecite{dmit02}, this system can be described by the Heisenberg model
\eqref{ham} with $J' = 1$ and $J'' = 1.05$. Although this configuration is outside of the
VBS region predicted by our bond-operator analysis, it is quite close to the VBS phase
boundary (see the orange diamond, Fig.~\ref{fig:phase-diag}).
It is conceivable that a combination of longer-range or ring-exchange interactions, which
arise in the proximity to the Mott transition,\cite{schmidt10} increase the level of
magnetic frustration, which is then released by a lattice dimerization through magnetoelastic
couplings (note that organic compounds of the \xdmit\ display a rather soft lattice).
As a result, a VBS phase can emerge, whose magnetic couplings are explicitly dimerized as
in our Hamiltonian \eqref{ham}.

Therefore we believe that the excitation spectrum of the organic compound may display
features qualitatively similar to the ones shown, e.g., in Fig.~\ref{fig:disp}~b, which
corresponds to a configuration ($J' = 1.5$, $J'' = 1$) close to the critical line. Hence
we predict the spin correlations in the VBS phase of \pdmit\ to be incommensurate, which
may be checked in future neutron scattering or NMR experiments. Moreover it would be
interesting to see whether one could drive the material into a state with magnetic LRO,
by moving farther away from the Mott transition (and thus reducing the influence of
magnetic couplings beyond nearest-neighbor exchange).


\section{Summary}

We have studied a dimerized Heisenberg AF on a spatially anisotropic
triangular lattice, with focus on its quantum paramagnetic (i.e. dimer) phase. Starting
from bond-operator mean-field theory, we have included interaction corrections to the
harmonic approximation at the Hartree-Fock level for the quartic term and in second-order
perturbation theory for the cubic term. We have shown that this Hartree-Fock-cubic
approximation gives sensible results for the triplon dispersion for the investigated part
of the parameter space (away from the one-dimensional limit). The resulting boundary of
the dimer phase where the triplon gap closes, Fig.~\ref{fig:phase-diag}, is one of our
main results; its location in the unfrustrated limit $J''=0$ is in good quantitative
agreement with QMC data.

The minimum wavevector $\bQ_0$ of the triplon spectrum displays a
commensurate-incommensurate transition as frustration is increased. At this CIT the
quadratic piece of the triplon dispersion vanishes in one
momentum direction. The closing
of the triplon gap at the CIT leads to a quantum Lifshitz point, the detailed study of
which is left for future work: it either leads to distinct power-law critical behavior or
it may even be inherently unstable, leading to a novel intermediate phase,
Fig.~\ref{fig:schem}. Remarkably, we find the critical $J'$ to be only slightly above
unity, suggesting that the origin of the non-trivial VBS phase seen in series-expansion
studies\cite{series-exp} for the non-dimerized model is indeed the enhanced fluctuations
near the quantum Lifshitz point. Finally, in the incommensurate regime of the dimer phase
the cubic triplon interaction leads to two-particle decay of triplons at high energies.

Our study paves the way to further investigations of frustrated dimerized magnets. As
those are difficult to access using QMC simulations, due to the inherent minus sign problem,
bond-operator as well as series-expansion techniques are often the methods of choice.
Clearly, the inclusion of longer-range and cyclic exchange interactions on triangular
and Kagome lattice geometries would be most interesting. On the methodological side, a
self-consistent treatment of the cubic term might further improve the numerical accuracy.
In addition, the formation of bound states of triplons -- which would signify the
instability towards phases other than those with magnetic LRO
-- should be studied; we leave this for future work. Finally, the interplay of
magnetoelastic couplings and longer-range exchange interactions should be studied near
the isotropic point, with an eye towards \pdmit.


\acknowledgments

We thank L. Fritz, M. Garst, and T. Vojta for helpful discussions.
This research was partially supported by the DFG through SFB 608
(K\"oln), GRK 1621 (Dresden), and FOR 960. R.L.D. also acknowledges
support by FAPESP (project No. 10/00479-6).


\appendix

\section{Classical phase diagram}
\label{ap:class}

To determine the classical phases of the model \eqref{ham}, we parametrize the spins
$\bS^1_i$ and $\bS^2_i$ (Fig.~\ref{fig:model}a) according to
\begin{eqnarray}
 \bS^1_i &=& \hat{e}_1\cos(\bQ\cdot\bR_i) + \hat{e}_2\sin(\bQ\cdot\bR_i),
\nonumber \\
&& \label{spins-class} \\
 \bS^2_i &=& \hat{e}_3\cos(\bQ\cdot\bR_i) + \hat{e}_4\sin(\bQ\cdot\bR_i),
\nonumber
\end{eqnarray}
which assumes coplanar order.
Here $\hat{e}_i$ are a set of unit vectors which obey:
$\hat{e}_1\cdot\hat{e}_2 = \hat{e}_3\cdot\hat{e}_4 = 0$,
$\hat{e}_1\cdot\hat{e}_3 = \hat{e}_2\cdot\hat{e}_4 = \cos\theta$,
$\hat{e}_2\cdot\hat{e}_3 = -\hat{e}_1\cdot\hat{e}_4 = \sin\theta$.
Substituting Eq.~\eqref{spins-class} into the Hamiltonian
\eqref{ham-underline}, it is easy to see that the total energy is
$E = (N/2)S^2J_\bQ(\theta)$ with
\begin{eqnarray}
 J_\bQ(\theta) &=& J'\cos\theta + 2J''\cos(Q_x) + \cos(Q_x + \theta)
\nonumber \\
     &+& \cos(\sqrt{3}Q_y + \theta) + \cos(Q_x + \sqrt{3}Q_y + \theta).
    \;\;
\label{class-energy}
\end{eqnarray}
For fixed $J'$ and $J''$, the ground state energy is determined
by minimizing Eq.~\eqref{class-energy} with respect to the components
of the vector $\bQ$ and the angle $\theta$.

\begin{figure}[!t]
\centerline{\includegraphics[width=6.7cm]{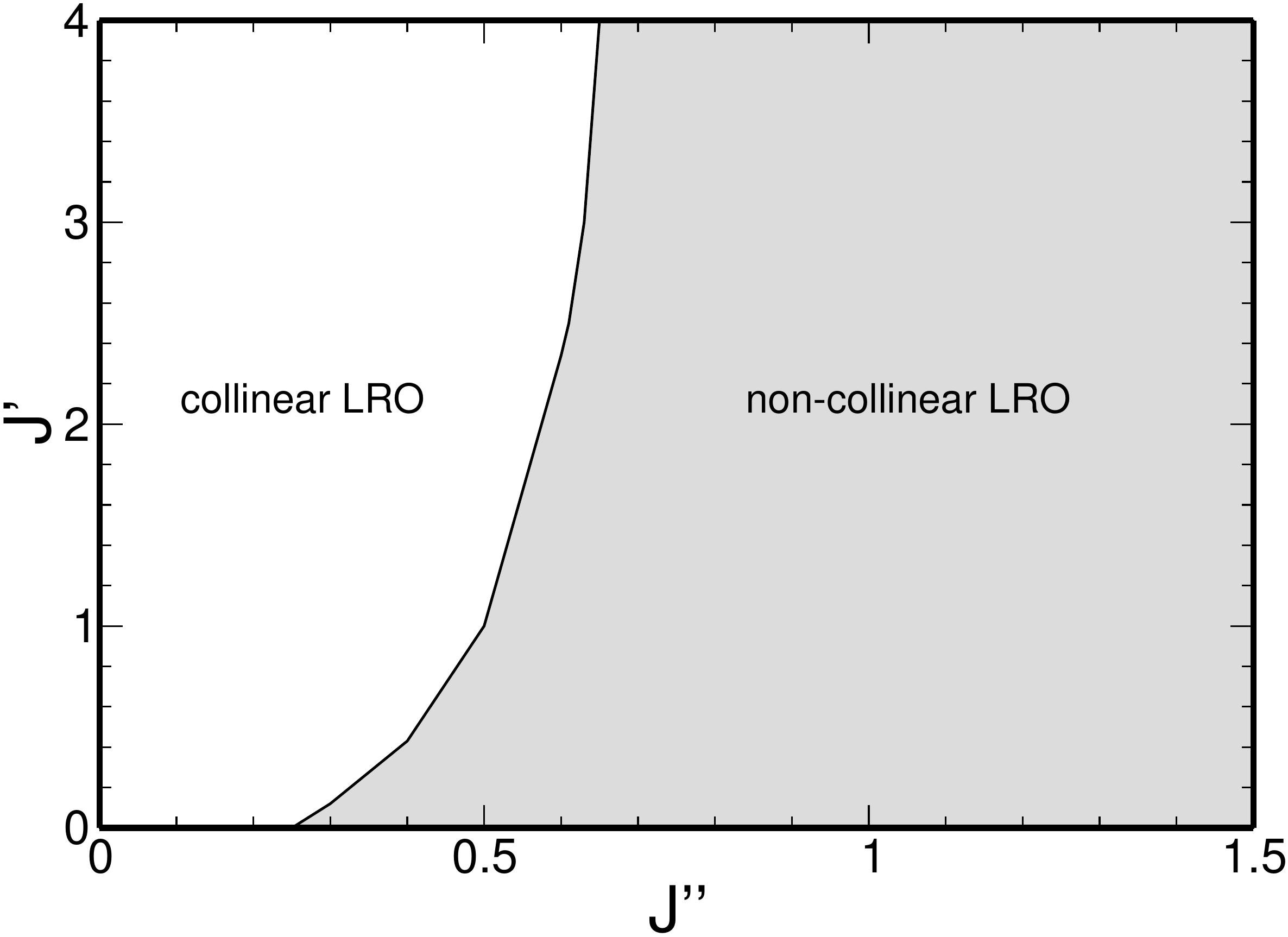}}
\caption{(Color online)
 Classical phase diagram for the Heisenberg model \eqref{ham}, obtained from
 numerically solving Eq.~\eqref{eqs-ordering}.
}
\label{fig:phase-diag-class}
\end{figure}

We find two phases, Fig.~\ref{fig:phase-diag-class}:
a) collinear order, with a commensurate ordering wavevector $\bQ=(0,0)$
and $\theta =\pi$, realized for $J'' \le J''_{\rm CIT}(J')$, and
b) non-collinear order, with incommensurate
$\bQ = (Q_{\rm xm},Q_{\rm ym})$ and
$\theta = \pi - Q_{\rm xm}/2 - \sqrt{3}Q_{\rm ym}$.
The components of $\bQ$ solve
\begin{eqnarray}
 &&\mu\sin(Q_x/2)\cos(\sqrt{3}Q_y) - \sin(Q_x/2 - \sqrt{3}Q_y) = 0,
\nonumber \\
&& \nonumber \\
&& 4J''\cos(Q_x/2) - \mu\cos(\sqrt{3}Q_y) = 1,
\label{eqs-ordering}
\end{eqnarray}
with $\mu = 2J'/(1 + J')$. Numerical solutions of \eqref{eqs-ordering} can be easily
obtained, which yield the phase boundary in Fig.~\ref{fig:phase-diag-class}.

Interestingly, it is possible to solve Eqs. \eqref{eqs-ordering} analytically
for the particular cases $J'= 0$, $1$, and $\infty$ which,
respectively, correspond to $\mu = 0$, $1$, and $2$. The solutions can
be written in the following way:
\begin{eqnarray}
Q_{\rm xm} &=& 2\arccos\left(\frac{1-\mu^2 + 2\mu}{4J'' - \mu(\mu-1)}\right),
\nonumber \\
&& \\
Q_{\rm ym} &=& \frac{\lambda_\mu}{\sqrt{3}}
               \arccos\left(\frac{1-\mu^2 + 2\mu}{4J'' - \mu(\mu-1)}\right).
\nonumber
\end{eqnarray}
Here, $\lambda_0 = -\lambda_2 = 1$ and $\lambda_1 = 0$.  Since the argument
of $\arccos(x)$ is within the range $-1 \le x \le 1$, we conclude that
the non-collinear phase is stable for $J''\ge J''_{\rm CIT}$ where
\begin{equation}
 J''_{\rm CIT} = \frac{1}{4} + \frac{J'}{2(1 + J')},
\label{eq-cit}
\end{equation}
valid for $J'= 0$, $1$, and $\infty$.
(For $J'= \infty$, we obtain $J''_{\rm CIT} = 0.75$ and $Q_{\rm xm,ym} = q_{\rm xm,ym}$,
i.e., the classical result matches the corresponding features of bond-operator result for
$S=1/2$.)

Quantum corrections to the classical phase boundary can be obtained using non-linear
spin-wave theory. For $J'=1$ this has recently been demonstrated to move the CIT boundary
to $J''=0.77$ at order $1/S$ (Ref.~\onlinecite{wang11}).


\section{Effective triplet Hamiltonian in real space}
\label{ap:details1}

Here we quote the full expressions for the real-space Hamiltonian
\eqref{ham-bond}, which can be obtained using straightforward algebra
via Eq.~\eqref{spin-bondop}.
\begin{eqnarray}
\mathcal{H}_0 &=& -\frac{3J'}{4}\sum_j s^\dagger_js_j,
\label{h0-real} \\
&& \nonumber \\
\mathcal{H}_2 &=& \frac{J'}{4}\sum_jt^\dagger_{j\,\alpha}t_{j\,\alpha}
        \nonumber \\
        &&  + \frac{1}{4} \sum_{j,n}g_2(n)\left[
            (s_js^\dagger_{j+n}t^\dagger_{j\,\alpha}t_{j+n\,\alpha} + {\rm H.c.})
            \right.
        \nonumber \\
        && + \left. (s^\dagger_js^\dagger_{j+n}t_{j\,\alpha}t_{j+n\,\alpha}
           + {\rm H.c.})\right],
\label{h2-real} \\
&& \nonumber \\
\mathcal{H}_3 &=& \frac{i}{4}\epsilon_{\alpha\beta\lambda}
        \sum_{j,n}\left\{ \left[
        (s^\dagger_jt_{j\,\alpha} + t^\dagger_{j\,\alpha}s_j)
         t^\dagger_{j+n\,\beta}t_{j+n\,\lambda}
        \right.\right.
\nonumber \\
    && \left.\left. + \,{\rm H.c.}\right] - \;(j \leftrightarrow  j+n) \right\},
\label{h3-real} \\
&& \nonumber \\
\mathcal{H}_4 &=& -\frac{1}{4}\epsilon_{\alpha\beta\lambda}\epsilon_{\alpha\mu\nu}
      \sum_{j,n}g_4(n)
      t^\dagger_{j\,\beta}t^\dagger_{j+n\,\mu}t_{j+n\,\nu}t_{j\,\lambda},
\label{h4-real}
\end{eqnarray}
where the $\sum_j$ runs over the sites of the dimerized lattice, and the
$g_m(n)$ coefficients read
\begin{eqnarray}
  g_2(n) &=& (2J'' - 1)\delta_{n,1} - \delta_{n,2} - \delta_{n,3},
\nonumber \\
  g_4(n) &=& (2J'' + 1)\delta_{n,1} + \delta_{n,2} + \delta_{n,3}.
\nonumber
\end{eqnarray}

\section{Self-consistent equations}
\label{ap:details2}

Here we derive the self-consistent
Eqs.~\eqref{self-mu}--\eqref{self-nzero} involved in the harmonic
approximation, and Eqs.~\eqref{self-mu-hf}--\eqref{self-nzero-hf}
related to the HF one. As mentioned in Sec.~\ref{sec:harmonic}, the
starting point is the saddle point conditions: (i) $\partial
\bar{E}_0/\partial N_0 = 0$ and (ii) $\partial \bar{E}_0/\partial \mu = 0$.

Let us consider the harmonic case. From the expression
\eqref{egs-harmonic} for the ground state energy, we have
\begin{equation}
  \frac{\partial \bar{E}_0}{\partial N_0} =
        -\frac{3}{8}J'N - \frac{1}{2}\mu N
        + \frac{3}{2}\sum_\bq \left(\frac{\partial\omega_\bq}{\partial N_0}
                                  - \frac{\partial A_\bq}{\partial N_0}  \right).
\label{eq-i}
\end{equation}
Since $\partial A_\bq/\partial N_0 = \partial B_\bq/\partial N_0 =
B_\bq/N_0$, see Eqs.~\eqref{coef-a}--\eqref{coef-b},
condition (i) yields Eq.~\eqref{self-mu}. Moreover,
\begin{equation}
  \frac{\partial \bar{E}_0}{\partial \mu} =
        -\frac{1}{2}N(N_0 - 1)
        + \frac{3}{2}\sum_\bq \left(\frac{\partial\omega_\bq}{\partial \mu}
                                  - \frac{\partial A_\bq}{\partial \mu}  \right).
\label{eq-ii}
\end{equation}
One can easily see that $\partial A_\bq/\partial \mu = -1$ and $\partial
\omega_\bq/\partial \mu = -A_\bq/\omega_\bq$, and therefore condition (ii)
leads to Eq.~\eqref{self-nzero}.

Turning to the Hartree-Fock approximation, the equivalent of
Eq.~\eqref{eq-i} is here given by
\begin{equation}
 \frac{\partial \bar{E}^{\rm HF}_0}{\partial N_0} =
        -\frac{3}{8}J'N - \frac{1}{2}\mu N
        + \frac{3}{2}\sum_\bq \left(\frac{\partial\bar{\omega}_\bq}{\partial N_0}
                                  - \frac{\partial\bar{A}_\bq}{\partial N_0}  \right).
\label{eq-iii}
\end{equation}
The contribution due to $E^{\rm HF}_0$, Eq.~\eqref{hf-coef}, can be
neglected since it is $\mathcal{O}(\bar{v}^2_\bq)$ and therefore small
compared to the other terms (see discussion in Sec.~\ref{sec:quartic}).
The last term of the above equation can be written as
\[
  I = \frac{3}{2}\sum_\bq
      \left( - 1 + \frac{\bar{A}_\bq}{\bar{\omega}_\bq} \right)
           \frac{\partial\bar{A}_\bq}{\partial N_0}
         - \frac{\bar{B}_\bq}{\bar{\omega}_\bq}\frac{\partial\bar{B}_\bq}{\partial N_0}
\]
with
\begin{eqnarray}
  \frac{\partial\bar{A}_\bq}{\partial N_0} &=&
    \frac{B_\bq}{N_0}
  - \frac{1}{N'}\sum_\bk\,\gamma_{\bk - \bq}
      \frac{\partial}{\partial N_0}\left( \frac{\bar{A}_\bk}{\bar{\omega}_\bk} \right),
\nonumber \\
  \frac{\partial\bar{B}_\bq}{\partial N_0} &=&
    \frac{B_\bq}{N_0}
  - \frac{1}{N'}\sum_\bk\,\gamma_{\bk - \bq}
      \frac{\partial}{\partial N_0}\left( \frac{\bar{B}_\bk}{\bar{\omega}_\bk} \right).
\nonumber
\end{eqnarray}
Assuming that
\begin{equation}
 \partial \bar{A}_\bq/\partial N_0 =
  \partial \bar{B}_\bq/\partial N_0 \approx B_\bq/N_0,
\label{eq-approx}
\end{equation}
and using the expressions \eqref{bogo-coef-hf}, we show that
$
  I \approx (3/N_0)\sum_\bq B_\bq \bar{v}_\bq
    \left(\bar{v}_\bq - \bar{u}_\bq \right),
$
and therefore Eq.~\eqref{self-mu-hf} follows from condition (i).
Some words about the assumption \eqref{eq-approx} are here in order:
we realize that it is a good approximation as long as $v_\bq$ is
small. Indeed, using the relations \eqref{bogo-coef-hf}, it is
possible to show that
\[
  \frac{\partial}{\partial N_0}\left(
   \frac{\bar{A}_\bk}{\bar{\omega}_\bk} \right) =
  \frac{4\bar{u}^2_\bk\bar{v}^2_\bk}{\bar{\omega}_\bk}
  \frac{\partial\bar{A}_\bq}{\partial N_0}
  +
  \frac{2\bar{u}_\bk\bar{v}_\bk}{\bar{\omega}_\bk}
  \left(\bar{v}^2_\bq + \bar{u}^2_\bq \right)
  \frac{\partial\bar{B}_\bk}{\partial N_0}.
\]
Substituting \eqref{eq-approx} in the above expression, one can easily
see that $\partial \bar{A}_\bq/\partial N_0
\approx B_\bq/N_0 + \mathcal{O}(\bar{v}_\bk)$.
Similar expression holds for $\partial
\bar{B}_\bq/\partial N_0$.

The analog of Eq.~\eqref{eq-ii} reads
\begin{equation}
  \frac{\partial \bar{E}^{\rm HF}_0}{\partial \mu} =
        -\frac{1}{2}N(N_0 - 1)
        + \frac{3}{2}\sum_\bq
           \left(\frac{\partial\bar{\omega}_\bq}{\partial \mu}
             - \frac{\partial\bar{A}_\bq}{\partial \mu}  \right).
\label{eq-iv}
\end{equation}
Here
\begin{eqnarray}
  \frac{\partial\bar{A}_\bq}{\partial \mu} &=&
    - 1 - \frac{1}{N'}\sum_\bk\,\gamma_{\bk - \bq}
      \frac{\partial}{\partial \mu}\left( \frac{\bar{A}_\bk}{\bar{\omega}_\bk} \right),
\nonumber \\
  \frac{\partial\bar{B}_\bq}{\partial \mu} &=&
  - \frac{1}{N'}\sum_\bk\,\gamma_{\bk - \bq}
      \frac{\partial}{\partial \mu}\left( \frac{\bar{B}_\bk}{\bar{\omega}_\bk} \right).
\nonumber
\end{eqnarray}
In order to be consistent with the above approximation, we assume that
$\partial\bar{A}_\bq/\partial \mu \approx -1$ and
$\partial\bar{B}_\bq/\partial \mu \approx 0$ which implies
$\partial\bar{\omega}_\bq/\partial \mu \approx
-\bar{A}_\bq/\bar{\omega}_\bq$. Notice that this is
indeed a reasonable approximation because condition (ii) yields to
nothing else but the conservation (on average) of the total
number of bosons per site of the dimerized lattice, i.e.,
Eq.~\eqref{self-nzero-hf}.



\begin{thebibliography}{}


\bibitem{review-subir}
S. Sachdev, in: {\sl Quantum magnetism},
Lecture Notes in Physics Vol. 645,
edited by U. Schollw\"ock, J. Richter, D. J. J. Farnell, and R. A. Bishop
(Springer, Berlin 2004);
S. Sachdev, Nature Physics {\bf 4}, 173 (2008).


\bibitem{misguich}
G. Misguich and C. Lhuillier, in {\sl Frustrated Spin Systems}, edited
by H. T. Diep (World Scientific, Singapore, 2004);
C. Lhuillier, preprint arXiv:cond-mat/0502464.



\bibitem{review-balents}
L. Balents, Nature {\bf 464}, 199 (2010).


\bibitem{rvb-triangle}
P. W. Anderson, Mater. Res. Bull. {\bf 8}, 153 (1973).



\bibitem{isotropic-model}
See, e.g.,
L. Capriotti, A. E. Trumper, and S. Sorella,
Phys. Rev. Lett. {\bf 82}, 3899 (1999);
W. Zheng, J. O. Fjarestad, R. R. P. Singh, R. H. McKenzie, and R. Coldea,
Phys. Rev. B {\bf 74}, 224420 (2006);
S. R. White and A. L. Chernyshev,
Phys. Rev. Lett. {\bf 99}, 127004 (2007).

\bibitem{schmidt10}
H.-Y.Yang, A. L\"auchli, F. Mila, and K. P. Schmidt,
Phys. Rev. Lett. {\bf 105}, 267204 (2010).


\bibitem{trumper99}
A. E. Trumper,
Phys. Rev. B {\bf 60}, 2987 (1999).


\bibitem{merino99}
J. Merino, R. H. McKenzie, J. B. Marston, and C. H. Chung,
J. Phys.: Condens. Matter {\bf 11}, 2965 (1999).


\bibitem{series-exp}
Z. Weihong, R. H. McKenzie, and R. R. P. Singh,
Phys. Rev. B {\bf 59}, 14367 (1999).


\bibitem{ccm}
R. F. Bishop, P. H. Y. Li, D. J. J. Farnell, and C. E. Campbell,
Phys. Rev. B {\bf 79}, 174405 (2009).


\bibitem{rg}
O. A. Starykh and L. Balents,
Phys. Rev. Lett. {\bf 98}, 077205 (2007).

\bibitem{kallin11}
S. Ghamari, C. Kallin, S.-S. Lee, and E. S. Sorensen,
Phys. Rev. B {\bf 84}, 174415 (2011).


\bibitem{weichsel11}
A. Weichselbaum and S. R. White,
Phys. Rev. B {\bf 84}, 245130 (2011).


\bibitem{spin-liquid}
S. Yunoki and S. Sorella,
Phys. Rev. B {\bf 74}, 014408 (2006);
D. Heidarian, S. Sorella, and F. Becca,
Phys. Rev. B {\bf 80}, 012404 (2009).

\bibitem{frg}
J. Reuther and R. Thomale,
Phys. Rev. B {\bf 83}, 024402 (2011).


\bibitem{xu09}
C. Xu and S. Sachdev,
Phys. Rev. B {\bf 79}, 064405 (2009).

\bibitem{hauke11}
P. Hauke, T. Roscilde, V. Murg, J I. Cirac, and R. Schmied,
New J. Phys. {\bf 13}, 075017 (2011).


\bibitem{cscucl}
R. Coldea, D. A. Tennant, A. M. Tsvelik, and Z. Tylczynski,
Phys. Rev. Lett. {\bf 86}, 1335 (2001).

\bibitem{cscubr}
T. Ono, H. Tanaka, H. Aruga Katori, F. Ishikawa, H. Mitamura, and T. Goto,
Phys. Rev. B {\bf 67}, 104431 (2003).

\bibitem{kappa-et}
Y. Shimizu, K. Miyagawa, K. Kanoda, M. Maesato, and G. Saito,
Phys. Rev. Lett. {\bf 91}, 107001 (2003);
Phys. Rev. B {\bf 73}, 140407(R) (2006).


\bibitem{dmit01}
M. Tamura, A. Nakao, and R. Kato,
J. Phys. Soc. Jpn. {\bf 75}, 093701 (2006).


\bibitem{dmit02}
Y. Shimizu, H. Akimoto, H. Tsujii, A. Tajima, and R. Kato,
Phys. Rev. Lett. {\bf 99}, 256403 (2007).


\bibitem{dmit03}
T. Itou, A. Oyamada, S. Maegawa, K. Kubo, H. M. Yamamoto, and R. Kato,
Phys. Rev. B {\bf 79}, 174517 (2009).

\bibitem{dmit_rev}
M. Tamura and R. Kato,
Sci. Technol. Adv. Mater. {\bf 10}, 024304 (2009).

\bibitem{dmit04}
T. Itou, A. Oyamada, S. Maegawa, M. Tamura, and R. Kato,
Phys. Rev. B {\bf 77}, 104413 (2008).

\bibitem{pattern}
The dimerization pattern in Fig.~\ref{fig:model} appears different from the one proposed
for \pdmit\ in Ref.~\onlinecite{dmit01}, but we note that the undistorted phase of
\pdmit\ is very close to the isotropic point with the three exchange couplings
being equal within 6\%, such that the dimer states are nearly degenerate.
One of our motivations to choose the dimer pattern of Fig.~\ref{fig:model} was to make
contact with the staggered dimerized square-lattice model of
Refs.~\onlinecite{wenzel08,stag-dimer}.

\bibitem{SachdevBhatt}
S. Sachdev and R. Bhatt, Phys. Rev. B {\bf 41}, 9323 (1990).


\bibitem{qlif1}
A. Dutta, B. K. Chakrabarti, and J. K. Bhattacharjee,
\prb {\bf 55}, 5619 (1997).


\bibitem{qlif2}
R. Ramazashvili,
\prb {\bf 60}, 7314 (1999).


\bibitem{kole06}
A. Kolezhuk and S. Sachdev,
\prl {\bf 96}, 087203 (2006).

\bibitem{zhito06}
M. E. Zhitomirsky, Phys. Rev. B {\bf 73}, 100404 (2006).


\bibitem{gia08}
T. Giamarchi, C. R\"uegg, and O. Tchernyshyov,
Nature Phys. {\bf 4}, 198 (2008).

\bibitem{wenzel08}
S. Wenzel, L. Bogacz, and W. Janke,
Phys. Rev. Lett. {\bf 101}, 127202 (2008).


\bibitem{stag-dimer}
L. Fritz, R. L. Doretto, S. Wessel, S. Wenzel, S. Burdin, and M. Vojta,
Phys. Rev. B {\bf 83}, 174416 (2011)


\bibitem{cherny09}
A. L. Chernyshev and M. E. Zhitomirsky,
Phys. Rev. B {\bf 79}, 144416 (2009).


\bibitem{chubukov95}
A. V. Chubukov and D. K. Morr,
Phys. Rev. B {\bf 52}, 3521 (1995).


\bibitem{kotov98}
V. N. Kotov, O. Sushkov, W. H. Zheng, and
J. Oitmaa, Phys. Rev. Lett. {\bf 80}, 5790 (1998).


\bibitem{kotov99}
V. N. Kotov, J. Oitmaa, O. P. Sushkov, and W. H. Zheng,
Phys. Rev. B {\bf 60}, 14613 (1999).

\bibitem{kotov99b}
P. V. Shevchenko, V. N. Kotov, and O. P. Sushkov,
Phys. Rev. B {\bf 60}, 3305 (1999).


\bibitem{comment01}
Strictly, the cubic interaction generates both normal and anomalous self-energies which
then yield a $2\times2$ matrix structure of the Dyson equation. However, in the $b$ basis, the
anomalous self-energy is small provided that $\bar{v}_\bk$ is small: For selected parameters,
we have checked that including the anomalous self-energy in the on-shell approximation
changes the triplon energy by less than 1\%.

\bibitem{wang11}
L. Wang and S. G. Chung,
preprint arXiv:1110.0377.


\end{thebibliography}
\end{document}